\documentclass[oneside]{article}
\usepackage[a4paper]{geometry}
\usepackage[latin1]{inputenc} 
\usepackage[T1]{fontenc} 
\usepackage{RR}
\usepackage{hyperref}
\usepackage{IEEEtrantools}
\usepackage{research5IEEE}
\usepackage{transparent }
\usepackage{mathrsfs,url,psfrag,graphicx,epsfig,mathtools,research5IEEE,amsmath,upgreek,
	amsfonts,amssymb,amsthm,tikz,array,wasysym,pifont,dsfont,balance,epstopdf}

\usetikzlibrary{arrows}

\newcommand{\GMACB}{\textnormal{G-MAC}$(b)$}
\newcommand{\GMACFB}{\textnormal{G-MAC-F}$(b)$}
\newcommand{\GMAC}{\textnormal{G-MAC}}
\newcommand{\GMACF}{\textnormal{G-MAC-F}}
\newtheorem{remark}{Remark}
\newcommand{\outage}{\textnormal{outage}}
\newcommand{\minr}{\textnormal{min}}

\newcommand{\ds}{\displaystyle}

\newcommand{\F}{ \mathrm{FB}}
\newcommand{\NF}{ \mathrm{NF}}
\newcommand{\FB}{ \mathrm{FB}}
\newtheorem{lemma}{Lemma}
\newtheorem{theorem}{Theorem}

\newtheorem{definition}{Definition}
\newtheorem{proposition}{Proposition}
\newtheorem{corollary}{Corollary}

\RRNo{8804}
\RRdate{November 2015}
\RRauthor{Selma Belhadj Amor \and  Samir M. Perlaza \and Ioannis Krikidis \and and \newline  H. Vincent Poor}
\authorhead{Belhadj Amor \& Perlaza \& Krikidis \&  Poor }
\RRtitle{L'utilisation de la voie de retour améliore la transmission simultanée d'information et d'énergie dans les canaux sans fils à accès multiple}
\RRetitle{Feedback Enhances Simultaneous Wireless Information and  Energy Transmission in Multiple Access Channels }
\titlehead{Simultaneous Wireless Information and  Energy Transmission in Multiple Access Channels}
\RRnote{
	Selma Belhadj Amor is with Laboratoire CITI (Joint Lab between Universit\'e de Lyon, INRIA, and INSA de Lyon), Lyon, France (selma.belhadj-amor@inria.fr).
	Samir M. Perlaza is with Laboratoire CITI (Joint Lab between Universit\'e de Lyon, INRIA and INSA de Lyon),  Lyon, France (samir.perlaza@inria.fr). He is also  with the Department of Electrical Engineering at Princeton University, Princeton, NJ 08544, USA. 
	Ioannis Krikidis is with the Department of Electrical and Computer Engineering at the University of Cyprus, Nicosia, Cyprus (krikidis@ucy.ac.cy).
	H. Vincent Poor is with the Department of Electrical Engineering at Princeton University, Princeton, NJ 08544, USA (poor@princeton.edu).
	Part of this work was presented in the Fifth International Conference on Communications and Networking (ComNet'2015), Hammamet, Tunisia, in November 2015 \cite{BelhadjAmor-COMNET-2015}. Part of this work was presented to the 2016 IEEE International Symposium on Information Theory (ISIT), Barcelona, Spain. This research is supported in part by the European Commission under Marie Sk\l{}odowska-Curie Individual Fellowship No.~659316, and in part by the U.S. National Science Foundation under Grants CNS-1456793 and ECCS-1343210.}

\RRresume{
Dans le présent-rapport, les limites fondamentales de la transmission simultanée d'information et d'énergie dans le canal Gaussien à accès multiple (G-MAC) avec et sans voie de retour sont déterminées.
L'ensemble des débits atteignables de transmission d'information et d'énergie (en bits par utilisation canal et en unités d'énergie par utilisation canal respectivement) est identifié.
En outre, on caractérise les limites fondamentales sur les débits individuels et le débit-somme de transmission de l'information pour un débit d'énergie donné à l'entrée d'un collecteur d'énergie
Dans le cas sans voie de retour, on démontre qu'un schéma d'atteignabilité, basé sur la division de puissance et sur l'annulation successive de l'interférence, est optimal.
En contrepartie, dans le cas avec voie de retour (G-MAC-F), un schéma d'atteignabilité, simple mais optimal, basé sur la division de puissance et sur le schéma d'Ozarow qui atteint la capacité, est présenté.  
Finalement, le gain en énergie induit par l'exploitation de la voie de retour est quantifié. La voie de retour peut au mieux dédoubler le débit d'énergie à fort rapport signal sur bruit (RSB) pour un débit-somme d'information égal à la capacité-somme. En revanche, l'utilisation de la voie de retour n'a aucun effect à très faibles RSBs.
}
\RRabstract{
In this report, the fundamental limits of simultaneous information and energy transmission in the two-user Gaussian multiple access channel (G-MAC) with and without feedback are fully characterized. More specifically, all the achievable information and energy transmission rates (in bits per channel use and energy-units per channel use, respectively) are identified. Furthermore, the fundamental limits on the individual and sum- rates given a minimum energy rate ensured at an energy harvester are also characterized. In the case without feedback, an achievability scheme based on power-splitting and successive interference cancellation is shown to be optimal. Alternatively, in the case with feedback (G-MAC-F), a simple yet optimal achievability scheme based on power-splitting and Ozarow's capacity achieving scheme is presented.  
Finally, the energy transmission enhancement induced by the use of feedback is quantified. Feedback can at most double the energy transmission rate at high SNRs when the information transmission sum-rate is kept fixed at the sum-capacity of the G-MAC,  but it has no effect at very low SNRs.
}
\RRmotcle{Voie de retour, canal Gaussien à accès multiple (G-MAC), transmission simultanée d'information et d'énergie, collecte d'énergie RF, région de capacité d'information-énergie. }
\RRkeyword{Feedback, Gaussian multiple access channel, simultaneous information and energy transmission, RF energy  harvesting, information-energy capacity region.}
\RRprojet{Socrate}  
\RCGrenoble 

\begin{document}
\makeRR   

\clearpage
\tableofcontents
\clearpage

\section{Introduction}
For decades, a traditional engineering perspective was to exclusively use radio frequency (RF) signals  for information transmission.  
However, a variety of modern wireless systems suggest that RF signals can be  simultaneously used for information and energy transmission~\cite{Bi15}. 
Typical examples of communications technologies already exploiting this principle are reported in \cite{PoWifi}.
Beyond the existing applications, simultaneous information and energy transmission (SEIT) appears as a promising technology for a variety of emerging applications including low-power short-range communication systems, sensor networks, machine-to-machine networks and body-area networks, among others~\cite{KRI1}. 

When a point-to-point communication involves sending energy along with information, it should be designed to simultaneously meet two goals: $(i)$ To reliably transmit information to a receiver at a given rate with a sufficiently small probability of error; and $(ii)$ To transmit energy to an energy harvester (EH) at a given rate with a sufficiently small probability of energy shortage. The EH might not necessarily be co-located with the information receiver. More specifically, the EH might possess a set of  antennas (rectennas) dedicated to  the energy harvesting task, which are independent of those dedicated to the information receiving task. 
In the special case in which the receiver and the EH are co-located, that is, they share the same antenna, a signal division via time-sharing or power-splitting must be implemented. In the former, a fraction of time the antenna is connected to the information receiver, whereas the remaining time it is connected to the EH. The latter implies a signal division in which part of the signal is sent to the information receiver and the remaining part is sent to the EH. This signal processing is out of the scope of this paper and the reader is referred to \cite{KRI1}. 
In the realm of information theory, the problem of point-to-point SEIT with a co-located EH is cast into a problem of information transmission subject to minimum energy constraints at the channel output \cite{VAR,Fouladgar-CL-2012}.  
From this perspective, the case with a co-located EH is a special case of the non-co-located EH case in which the input signal to the receiver is identical to the signal input to the EH. In this paper, the analysis of SEIT is general and focuses on the case of non-co-located EHs.
Information and energy transmission are often conflicting tasks, and thus subject to a trade-off between the information transmission rate (bits per channel use) and the energy transmission rate (energy-units per channel use). This trade-off is evidenced in finite constellation schemes, as highlighted in Popovski \textit{et al.}'s~\cite{POPOVSKI}. Consider the noiseless transmission of a $4$-PAM signal over a point-to-point channel with input alphabet $\{-2,-1,1,2\}$ and  with a  co-located EH. Given that the symbols $-2$ and $2$ (resp.~$-1$ and $1$)  deliver $4$ (resp.~$1$) energy-units/ch.use, without any energy rate constraint, the system conveys  a maximum of 2 bits/ch.use and $\frac52$ energy-units/ch.use by choosing all available symbols with equal probability. However, if
the received energy rate must be for instance at least $4$ energy-units/ch.use, the maximum information rate is 1 bit/ch.use. This is mainly because the transmitter is forced to communicate using only  the symbols capable of delivering the maximum energy rate.
From this simple example, it is easy to see how additional energy rate constraints may hinder information transmission in a point-to-point scenario.

In a multi-user scenario, the information-energy rate trade-off is more involved. Usually, users must coordinate their transmission strategies and cooperate so as to achieve the energy rate requirement. Consider for instance a network in which one single transmitter simultaneously transmits energy to an EH  and information to an information receiver. Assume that this transmitter is required to deliver an energy rate that is less than what it is able to deliver by only transmitting information. In this case, such a transmitter is able to fulfill the energy-transmission task independently of the behavior of the other transmitters. More importantly, it can use all its available power budget to maximize its information transmission rate while still being able of meeting the energy rate constraint. In this case, the minimum energy rate constraint does not play a fundamental role.
On the other hand, when the same transmitter is requested to deliver an energy rate that is higher than what it is able to deliver by only transmitting information, its behavior is totally dependent on the behavior of the other transmitters. Indeed, it depends on whether or not other transmitters are transmitting signals using an average
power such that the energy rate is met. In this case, the minimum energy rate constraint drastically affects the way that the transmitters interact with each other. More critical scenarios are the cases in which the requested energy rate is less than what all transmitters are able to deliver by simultaneously transmitting information using all the available individual power budgets.  In these cases, none of the transmitters can unilaterally ensure reliable energy transmission at the requested rate. Hence, transmitters must engage in a mechanism through which an energy rate that is higher than the energy delivered by exclusively transmitting information-carrying signals is ensured at the EH. 
This suggests, for instance, sending signals with correlation to increase the received energy rates.  This correlation can result from the use of power splits in which the transmitted symbols are formed by an information-carrying and an energy-carrying component. The latter typically consists in signals that are known at all devices and can be constructed such that the energy captured at the EH is maximized.

Most of the existing  studies of  SEIT follow  a signal-processing or networking approach and focus mainly on the feasibility aspects. For instance, optimization of beamforming strategies  was considered for multi-antenna broadcast channels in~\cite{HUANGLARSSON,XIANGTAO}, and \cite{ZHANGHO}, and  for multi-antenna interference channels in~\cite{PARKCLERCKX}. %
SEIT was also studied in the general realm of cellular systems in \cite{HUANGLAU} as well as in multi-hop relaying systems in  \cite{Fouladgar-CL-2012,Ding14,GOYU,ISHIBASHI,LUOZHANGLETAIEF}, and \cite{NASIRZHOU}. 
Other studies in the two-way channel are reported in \cite{POPOVSKI} and  in graphical unicast and multicast networks in~\cite{FOULADGARSIMEONE13}. 

From an information-theoretic viewpoint, the pioneering works by Varshney in~\cite{VAR} and~\cite{VAR12}, as well as Grover and Sahai in~\cite{GRO} provided the fundamental limits on SEIT in  point-to-point channels with co-located EH. More specifically,  the case of  the single-link point-to-point channel was discussed in~\cite{VAR} while  the case of parallel-links point-to-point channel was studied in~\cite{VAR12} and \cite{GRO}. 
Despite the vast existing literature on this subject, the fundamental limits of SEIT are still unknown in most multi-user channels.  Multi-hop and multi-antenna wiretap channels under minimum received energy rate constraints were considered in~\cite{Fouladgar-CL-2012} and \cite{BenawanUlukus2014}, respectively. 
In the case of the discrete memoryless multiple access channel (DM-MAC), the trade-off between information rate and energy rate has been studied in~\cite{Fouladgar-CL-2012}. Therein, Fouladgar {\it et al.} characterized the information-energy capacity region of the two-user DM-MAC when a minimum energy rate is required at the input of the receiver (the receiver and the EH are co-located). 
An extension of the work in~\cite{Fouladgar-CL-2012} to the Gaussian multiple access channel (G-MAC) is far from trivial due to the fact that the information-energy capacity region involves an auxiliary random-variable that cannot be eliminated as in the case without energy constraints.
Moreover, different energy rate constraints for the G-MAC have also been  investigated. For instance, Gastpar~\cite{GASTPAR04} considered the G-MAC under a maximum received energy rate constraint. Under this assumption, channel-output feedback has been shown not to increase the information capacity region. 
More generally, the use of feedback in the $K$-user G-MAC, even without energy rate constraints, has been shown to be of limited impact in terms of information sum-rate improvement. This holds even in the case of perfect feedback. More specifically, feedback  increases the information sum-capacity in the G-MAC by at most $\frac{\log_2(K)}{2}$ bits per channel use~\cite{Kramer02}. Hence, the use of feedback is difficult to justify from the point of view of exclusively transmitting  information. 

\subsection{Contributions}

This paper studies the fundamental limits of SEIT in the two-user G-MAC with an EH, with and without feedback. It shows that when the goal is to simultaneously transmit both information and energy, feedback can significantly improve the global performance of the system in terms of both information and energy transmission rates. 
More specifically,  the paper provides the first full characterization of the information-energy capacity region for the G-MAC with and without feedback, i.e., all the achievable information and energy transmission rates in bits per channel use and energy-units per channel use, respectively. Furthermore, the fundamental limits on the individual and sum- rates given a minimum energy rate ensured at the EH are also provided. 
In the case without feedback, an achievability scheme based on power-splitting and successive interference cancellation is shown to be optimal. Alternatively, in the case with feedback (G-MAC-F), a simple yet optimal achievability scheme that is based on power-splitting and Ozarow's capacity achieving scheme is presented. 
Although the proofs of achievability and converse build upon standard information-theoretic techniques, extending these techniques to account for the energy constraint involves many challenges. For instance, to derive upper bounds on the achievable information-energy rate triplets, there are two parts to consider: one that is related to the information transmission for which Fano's inequality is used, and another that is related to the energy transmission for which  concentration inequalities are used to derive an upper bound on the energy rate. 
Finally, the enhancement of the energy transmission rate induced by the use of feedback is quantified.  It is shown that feedback can at most double the energy transmission rate at high SNRs when the information transmission sum-rate is kept fixed at the sum-capacity of the G-MAC,  but it has no effect at very low SNRs.

\subsection{Organization of the Report}
The remainder of the report is structured as follows. Sec.~\ref{SecProblemFormulation} formulates the problem of SEIT in the two-user \GMACF~and \GMAC~with a non-co-located EH.
Secs.~\ref{SecMainResultsCapacity}-\ref{SecMainResultsEnergy} show the main results of this paper for the \GMAC~and the \GMACF~with an EH. 
Namely, for both settings the following fundamental limits are derived:   
$(a)$ the  information-energy capacity region; and $(b)$ the maximum information individual rates and sum-rates that can be achieved given a targeted energy rate. A global comparison of the fundamental limits in terms of information transmission rates is provided in Sec.~\ref{SecMainResultsComments}.
In Sec.~\ref{SecMainResultsEnergy}, the maximum energy rate  improvement that can be obtained at the input of the EH  by using feedback given a targeted information rate is characterized as well as its low and high SNR asymptotics.  
Finally, Sec.~\ref{SecConclusion} concludes the report and discusses possible extensions.  The appendices expose the proofs of the main results.
\section{Gaussian Multiple Access Channel With Feedback and Energy Harvester}\label{SecProblemFormulation}
\begin{figure}[t]
	\centering{
		\includegraphics{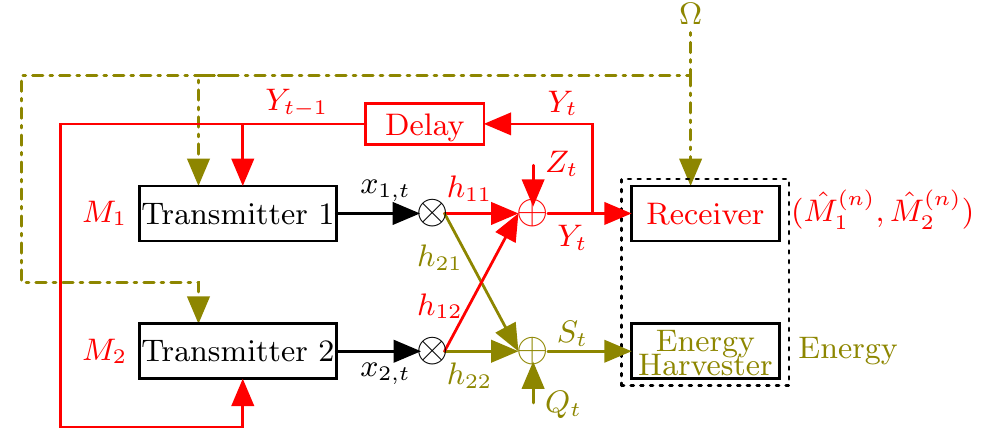}
	}
	\caption{Two-user memoryless G-MAC-F with an EH.}
	\label{FigMAC-FB}
\end{figure}

\begin{figure}[t]
	\centering{
		\includegraphics{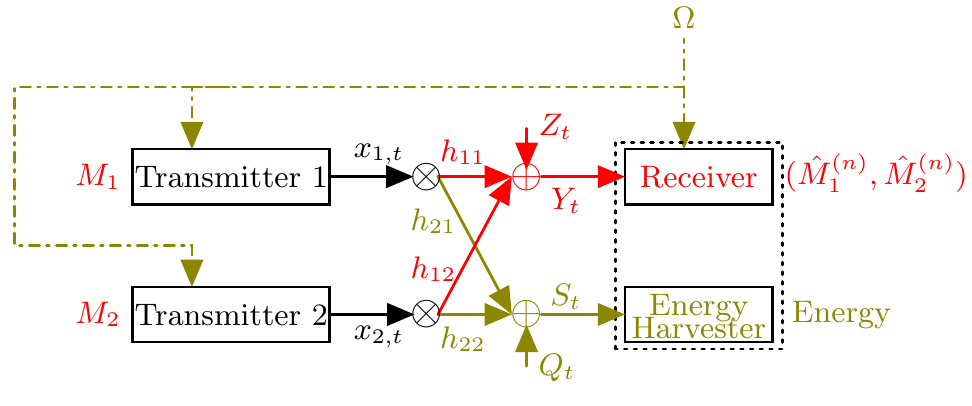}
	}
	\caption{Two-user memoryless G-MAC with an EH.}
	\label{FigMAC}
\end{figure}

Consider the two-user memoryless G-MAC with an EH with perfect channel-output-feedback  (G-MAC-F) in Fig.~\ref{FigMAC-FB} and without feedback in Fig.~\ref{FigMAC}.
In both channels, at each channel use $t\in \mathbb{N}$, $X_{1,t}$ and $X_{2,t}$ denote the real symbols sent by transmitters~1 and~2, respectively. Let $n\in\mathbb{N}$ denote the blocklength.
The receiver observes the real channel output 
\begin{equation}
\label{EqY}
Y_{1,t}=h_{11} X_{1,t} + h_{12} X_{2,t} + Z_t,
\end{equation}
and the EH observes 
\begin{equation}
\label{EqS}
Y_{2,t} = h_{21} X_{1,t} + h_{22} X_{2,t} + Q_t,
\end{equation}
where $h_{1i}$ and $h_{2i}$ are the corresponding constant non-negative real channel coefficients from transmitter $i$ to the receiver and the EH, respectively.
The channel coefficients are assumed to satisfy the following $\set{L}_2$-norm condition:
\begin{equation} 
\forall j\in \{1,2\},\quad \|\vect{h}_j\|^2 \leqslant 1,
\end{equation} 
with $\vect{h}_j\eqdef \trans{(h_{j1},h_{j2})}$ to satisfy the principle of conservation of energy. 

The noise terms $Z_{t}$ and $Q_{t}$ are realizations of two identically distributed 
zero-mean unit-variance real Gaussian random variables. In the following,  there is no particular assumption on the joint distribution of $Q_t$ and $Z_t$.

In the G-MAG-F with an EH, a perfect feedback link from the receiver to transmitter~$i$ allows at the end of each channel use~$t$, the observation of the channel output $Y_{t-d}$ at transmitter~$i$, with $d \in \mathbb{N}$ the delay of the feedback channel. Without any loss of generality, the delay is assumed to be the same from the receiver to both transmitters and equivalent to one channel use, i.e., $d = 1$.

Within this context, two main tasks are to be simultaneously accomplished: information transmission and energy transmission.

\subsection{Information Transmission}
The goal of the communication is to  convey the independent messages $M_1$ and $M_2$ from transmitters~1 and~2 to the common receiver.  The messages $M_1$ and $M_2$ are independent of the noise terms $Z_{1}, \ldots, Z_{n}$,  $Q_{1}, \ldots, Q_{n}$ and uniformly distributed over the sets $\set{M}_1\triangleq\{1,\dots, \lfloor 2^{n R_1}\rfloor\}$ and $\set{M}_2\triangleq\{1,\dots, \lfloor 2^{n R_2}\rfloor\}$, where $R_1$ and $R_2$ denote the information transmission rates and $n\in \mathbb{N}$ the blocklength. 

In the G-MAC-F with an EH, at each time $t$,  the existence of feedback links allows the $t$-th symbol of transmitter~$i$ to be dependent on all previous channel outputs $Y_1,\dots,Y_{t-1}$ as well as its message index $M_i$ and a randomly generated index $\Omega \in \{1,\dots, \lfloor 2^{n R_{r}}\rfloor\}$, with $R_r \geqslant 0$. The index $\Omega$ is independent of both $M_1$ and $M_2$ and assumed to be known by all transmitters and the receiver. More specifically,
\begin{subequations}
	\label{eq:channelinputs}
	\begin{IEEEeqnarray}{rCL}
		X_{i,1}&=&f_{i,1}^{(n)} (M_i,\Omega)\quad\text{and}\\
		X_{i,t}&=&f_{i,t}^{(n)} (M_i,\Omega, Y_{1,1},\dots,Y_{1,t-1}),\quad t\in\{2,\dots,n\},\IEEEeqnarraynumspace
	\end{IEEEeqnarray}
\end{subequations}
for some encoding functions 
\begin{IEEEeqnarray}{rCL}
	f_{i,1}^{(n)} & \colon & \set{M}_i\times \mathbb{N}  \to \mathbb{R} \quad\text{ and } \\
	f_{i,t}^{(n)}  & \colon & \set{M}_i \times \mathbb{N} \times \mathbb{R}^{t-1}\to \mathbb{R}. 
	\label{eq:mac_encoding}
\end{IEEEeqnarray}

In the G-MAC with an EH, at each time $t$,  the $t$-th symbol of transmitter~$i$ is
\begin{subequations}
	\begin{IEEEeqnarray}{rCL}
		X_{i,t}&=&g_{i,t}^{(n)} (M_i,\Omega),\quad t\in\{1,\dots,n\},
	\end{IEEEeqnarray}
\end{subequations}
where  $g_{i,t}^{(n)}\colon  \set{M}_i \times \mathbb{N} \to \mathbb{R}$ is the encoding function.

In the G-MAC-F and in the G-MAC with an EH, for all $i\in \{1,2\}$, transmitter~$i$'s channel inputs $X_{i,1}, \ldots, X_{i,n}$ satisfy an expected average \emph{input power constraint}
\begin{IEEEeqnarray}{C}
	\label{EqPowerConstraint}
	\frac1n \sum_{t=1}^n \E{X_{i,t}^2} \leqslant P_i,
\end{IEEEeqnarray}
where $P_i$ denotes the average transmit power of transmitter~$i$ in energy-units per channel use and where the expectation is over the message indices, the random index, and the noise realizations prior to channel use $t$. The dependence of $X_{i,t}$ on $Y_{1,1},\dots,Y_{1,t-1}$ (and thus on $Z_{1},\dots,Z_{t-1}$) is shown by \eqref{eq:channelinputs}.

The G-MAC-F and G-MAC with an EH are fully described by the signal to noise ratios (SNRs): $\SNR_{ji}$, with $\forall (i,j) \in \{1,2\}^2$. These SNRs are defined as follows
\begin{equation}
\label{eq:SNR}
\SNR_{ji}\eqdef |h_{ji}|^2 P_{i},
\end{equation}
given the normalization over the noise powers. 
The receiver produces an estimate $(\hat{M}_1^{(n)},\hat{M}_2^{(n)})=\Phi^{(n)}(Y_{1,1},\dots,Y_{1,n},\Omega)$ of the message-pair $(M_1,M_2)$ via a decoding function $\Phi^{(n)}\colon \mathbb{R}^{n}\times \mathbb{N}\to \set{M}_1\times \set{M}_2$, 
and the average probability of error is 
\begin{IEEEeqnarray}{rCL} 
	P_{\error}^{(n)}(R_1,R_2)&\eqdef&\Pr\left\{ (\hat{M}_1^{(n)},\hat{M}_2^{(n)}) \neq (M_1,M_2)\right\}.
\end{IEEEeqnarray} 

%
\subsection{Energy Transmission}
Let $b\geqslant 0$ denote the minimum energy rate that must be guaranteed at the input of the EH in the G-MAC-F. This rate $b$ (in energy-units per channel use) must satisfy
\begin{equation}
\label{EnergyCst}
0 \leqslant  b \leqslant  1 + \SNR_{21} + \SNR_{22} + 2 \sqrt{\SNR_{21}\SNR_{22}}, 
\end{equation}
for the problem to be feasible. In fact,  $1 + \SNR_{21} + \SNR_{22} + 2 \sqrt{\SNR_{21}\SNR_{22}}$ is the maximum energy rate that can be achieved at the input of the EH given the input power constraints in~\eqref{EqPowerConstraint}. This rate can be  achieved when the transmitters use all their power budgets to send fully correlated channel inputs.

The  empirical energy transmission rate (in energy-units per channel use) induced by the sequence $(Y_{2,1},\dots,Y_{2,n})$ at the input of the EH  is
\begin{equation}\label{EqEH}
B^{(n)}\eqdef \frac{1}{n}\sum_{t = 1}^{n} Y_{2,t}^2.
\end{equation}

The goal of the energy transmission is to guarantee that the empirical energy rate $B^{(n)}$ is not less than a given operational energy transmission rate $B$ that must satisfy 
\begin{equation}
\label{EqB}
b \leqslant B \leqslant 1 + \SNR_{21} +\SNR_{22} + 2\sqrt{\SNR_{21} \SNR_{22}}.
\end{equation}
Hence, the probability of energy outage is defined as follows:
\begin{IEEEeqnarray}{rCl}
	P_\outage^{(n)}(B)&\eqdef&\Pr\left\{B^{(n)}<B-\epsilon\right\},
\end{IEEEeqnarray}
for some $\epsilon>0$ arbitrarily small.

Note that $b$ denotes the minimum tolerable energy rate, whereas $B$ denotes the operating energy rate.

In the sequel, for ease of notation, the acronyms  \GMACFB~and \GMACB~refer to the \GMACF~and the \GMAC~with an EH depicted in Fig.~\ref{FigMAC-FB} and Fig.~\ref{FigMAC}, respectively, with fixed  SNRs:  $\SNR_{11}$, $\SNR_{12}$, $\SNR_{21}$, and $\SNR_{22}$,  and  minimum energy rate constraint $b$ at the input of the EH.

\subsection{Simultaneous Energy and Information Transmission (SEIT)}
The \GMACFB~(and \GMACB, respectively) 
is said to operate at the information-energy rate triplet $(R_1, R_2, B) \in [0,\infty)\times [0,\infty)  \times [b, \infty)$ when both transmitters and the receiver use a transmit-receive configuration such that:
$(i)$~reliable communication at information rates $R_1$ and $R_2$ is ensured; and 
$(ii)$ the empirical energy transmission rate  in~\eqref{EqEH}  at the input of the EH during the entire blocklength is not lower than~$B$. A formal definition is given below.
\begin{definition}[Achievable Rates]\label{DefAchievableTriples}
	The triplet $(R_1, R_2, B) \in [0,\infty)\times [0,\infty) \times [b, \infty)$ is achievable in the \GMACF$(b)$ (and \GMAC$(b)$, resp.) if there exists a sequence of encoding and decoding functions  $\big\{\{f_{1,t}^{(n)}\}_{t=1}^n,\{f_{2,t}^{(n)}\}_{t=1}^n,\Phi^{(n)}\big\}_{n=1}^\infty$ (and $\big\{\{g_{1,t}^{(n)}\}_{t=1}^n,\{g_{2,t}^{(n)}\}_{t=1}^n,\Phi^{(n)}\big\}_{n=1}^\infty$, resp.) such that both the average error probability  and the energy-outage probability tend to zero as the blocklength $n$ tends to infinity. That is,
	\begin{IEEEeqnarray}{lcl}
		\label{EqProbError}
		\limsup_{n \rightarrow \infty}\;  P_{\error}^{(n)} (R_1,R_2) & = & 0,\\
		\label{EqProbPower}
		\limsup_{n \rightarrow \infty}\;  P_\outage^{(n)}(B)  & = & 0\quad\text{ for any $\epsilon>0$.}
	\end{IEEEeqnarray}
\end{definition}
Often, increasing the energy transmission rate implies decreasing the information transmission rates and \emph{vice-versa}.
This trade-off is accurately captured by the notion of  \emph{information-energy capacity region}. 
\begin{definition}[Information-Energy Capacity Region]\label{DefCER}
	The information-energy capacity region of the \GMACF$(b)$ (and \GMAC$(b)$, resp.), denoted by $\set{E}_b^\FB (\SNR_{11},\SNR_{12},\SNR_{21},\SNR_{22})$ ($\set{E}_b (\SNR_{11},\SNR_{12},\SNR_{21},\SNR_{22})$, resp.) is the closure of all achievable information-energy rate triplets  $(R_1, R_2, B)$.
\end{definition}

\section{Information-Energy Capacity Region} 
\label{SecMainResultsCapacity}
For any non-negative SNRs: $\SNR_{11}$, $\SNR_{12}$, $\SNR_{21}$, and $\SNR_{22}$,  and for any minimum energy rate constraint $b$  
satisfying \eqref{EnergyCst}, the main results presented in this report are provided in terms of the information-energy capacity region (Def.~\ref{DefCER}). 
The results for the \GMACB~are  a particularization  of the results for the \GMACFB. The interest of presenting these results separately stems from the need for comparing both cases.

\subsection{Case With Feedback}
\label{SecMainResultsCapacityFB}

The information-energy capacity region of the \GMACFB~is fully characterized by the following theorem.
\begin{theorem}[Information-Energy Capacity Region of the \GMACFB] 
	\label{Theorem-EC-Region-MAC-F}
	The information-energy capacity region $\set{E}_b^\FB \left(\SNR_{11},\SNR_{12},\SNR_{21},\SNR_{22} \right)$ of the \GMACFB~is the set of information-energy rate  triplets $(R_1, R_2, B)$ that satisfy\begin{subequations}\label{eq:regthm1}	
		\begin{IEEEeqnarray}{ccccl}
			\label{EqEC1}0&\leqslant & R_1 & \leqslant & \frac{1}{2} \log_2\left( 1 + \beta_1 \;\SNR_{11}\left( 1 - \rho^2 \right)  \right),\\
			\label{EqEC2}
			0&\leqslant &R_2 & \leqslant & \frac{1}{2} \log_2\left( 1 + \beta_2\; \SNR_{12} \left( 1 - \rho^2 \right)   \right),\\
			\label{EqEC3}
			0&\leqslant & R_1 + R_2 & \leqslant & \frac{1}{2} \log_2 \big( 1 +\beta_1\;\SNR_{11} + \beta_2\;\SNR_{12} + 2 \rho  \sqrt{\beta_1 \SNR_{11}\beta_2 \SNR_{12}} \big), \IEEEeqnarraynumspace\\
			\label{EqEC4}
			b&\leqslant &B  & \leqslant &  1 + \SNR_{21} + \SNR_{22} + 2 \rho \sqrt{\beta_1 \SNR_{21} \beta_2 \SNR_{22}}\nonumber\\&&&& +2 \sqrt{(1-\beta_1)\SNR_{21}\;(1-\beta_2)\SNR_{22}},
		\end{IEEEeqnarray}
	\end{subequations}
	with $(\rho,\beta_1,\beta_2) \in \left[ 0, 1 \right]^3 $.
\end{theorem}
\begin{IEEEproof}
	The proof of Theorem~\ref{Theorem-EC-Region-MAC-F} is presented in Appendix~\ref{SecProofThm3}.
\end{IEEEproof}
\subsection{Case Without Feedback}
\label{SecMainResultsCapacityNF}

The information-energy capacity region of the \GMACB~is fully characterized by the following theorem.
\begin{theorem}[Information-Energy Capacity Region of the \GMACB] 
	\label{Theorem-EC-Region-MAC-NF}
	The information-energy capacity region $\set{E}_b\left(\SNR_{11},\SNR_{12},\SNR_{21},\SNR_{22} \right)$ of the \GMACB~is the set of all  information-energy rate triplets $(R_1, R_2, B)$ that satisfy
	\begin{subequations}
		\label{eq:regprop1}	
		\begin{IEEEeqnarray}{ccccl}
			0&\leqslant & R_1 & \leqslant & \frac{1}{2} \log_2\left( 1 + \beta_1 \;\SNR_{11}  \right)\label{eq:thm1_c1},\\
			0&\leqslant & R_2 & \leqslant & \frac{1}{2} \log_2\left( 1 + \beta_2\; \SNR_{12} \right)\label{eq:thm1_c2},\\
			0&\leqslant & R_1 + R_2 & \leqslant & \frac{1}{2} \log_2 \big( 1 +\beta_1\;\SNR_{11} + \beta_2\;\SNR_{12}\big), \IEEEeqnarraynumspace\label{eq:thm1_c12}\\
			b&\leqslant &B  & \leqslant &  1 + \SNR_{21} + \SNR_{22}  + 2  \sqrt{(1-\beta_1)\SNR_{21} (1-\beta_2)\SNR_{22}},\label{eq:thm1_e}
		\end{IEEEeqnarray}
	\end{subequations}
	with $(\beta_1,\beta_2) \in \left[ 0, 1 \right]^2 $.
\end{theorem}
\begin{IEEEproof}The proof of Theorem~\ref{Theorem-EC-Region-MAC-NF} is presented in Appendix~\ref{ProofTheorem-EC-Region-MAC-NF}.
\end{IEEEproof}

\begin{remark}
	For any non-negative $\SNR_{11}$, $\SNR_{12}$, $\SNR_{21}$, and $\SNR_{22}$,  and for any $b$ satisfying \eqref{EnergyCst},  the information-energy capacity region of the \GMAC$(b)$ is included in the information-energy capacity region of the \GMACF$(b)$, i.e.,
	\begin{equation}
	\set{E}_b \left(\SNR_{11},\SNR_{12},\SNR_{21},\SNR_{22}\right) \subseteq \set{E}_b^{\FB} \left(\SNR_{11},\SNR_{12},\SNR_{21},\SNR_{22}\right).
	\end{equation}
	Note that this inclusion can be strict. For instance, any rate triplet $(R_1, R_2, B)$ that is achievable in the \GMACF$(b)$, for a given minimum energy constraint $b$, and  for which $R_1+R_2$ equals the perfect feedback sum-capacity cannot be achieved in the \GMAC$(b)$.
	Note also that if $b = 1 + \SNR_{21} + \SNR_{22} + 2\sqrt{\SNR_{21}\SNR_{21}}$, then both information-energy capacity regions are equal as they only contain the point $(0,0,b)$.
\end{remark}

The remainder of this section highlights some important observations on the achievability and converse proofs of Theorem~\ref{Theorem-EC-Region-MAC-F} and Theorem~\ref{Theorem-EC-Region-MAC-NF}. The corresponding proofs are presented in Appendix~\ref{SecProofThm3} and Appendix~\ref{ProofTheorem-EC-Region-MAC-NF}, respectively.
\subsection{Comments on the Achievability}
The achievability scheme in the proof of Theorem~\ref{Theorem-EC-Region-MAC-F} is based on power-splitting and Ozarow's capacity-achieving scheme~\cite{Ozarow-TIT-1984}.
From an achievability standpoint, the parameters $\beta_1$ and $\beta_2$ in Theorem~\ref{Theorem-EC-Region-MAC-F} might be interpreted as the fractions of average power that transmitters $1$ and~$2$ allocate for information transmission. More specifically, transmitter $i$ generates two signals: an information-carrying (IC) signal with average power $\beta_i P_i$ energy-units per channel use; and a no-information-carrying (NIC) signal with power $(1-\beta_i) P_i$ energy-units per channel use. The IC signal is constructed using Ozarow's scheme~\cite{Ozarow-TIT-1984}. The role of the NIC signal is to exclusively transmit energy from the transmitter to the EH. Conversely, the role of the IC signal is twofold:  information transmission from the transmitter to the receiver and energy transmission from the transmitter to the EH.

The parameter $\rho$ is the average Pearson correlation coefficient between the IC signals sent by both transmitters. This parameter plays a fundamental role in both information transmission and energy transmission. 
Note for instance that the upper-bounds on the information sum-rate~\eqref{EqEC3} and on the energy harvested per unit-time~\eqref{EqEC4} monotonically increase with $\rho$, whereas the upper-bounds on the individual rates~\eqref{EqEC1} and~\eqref{EqEC2} monotonically decrease with~$\rho$.
If $\beta_1\neq 0$ and $\beta_2\neq 0$,  let $\rho^\star(\beta_1,\beta_2)$ be the unique solution in $(0,1)$ to the  following equation in $\rho$:
\begin{IEEEeqnarray}{lcl}
	\label{eq:rhostar}
	\lefteqn{1+\beta_1\;\SNR_{11}+\beta_2\;\SNR_{12}+2\rho  \sqrt{\beta_1\SNR_{11}\beta_2 \SNR_{12} } }\nonumber\qquad\qquad\\
	& =& \left(1+\beta_1\;\SNR_{11} (1- \rho^2)\right) \left(1+\beta_2\;\SNR_{12} (1- \rho^2)\right),\IEEEeqnarraynumspace 
\end{IEEEeqnarray}
otherwise, let  $\rho^\star(\beta_1,\beta_2)=0$. When $\rho=\rho^\star(\beta_1,\beta_2)$, the sum of~\eqref{EqEC1} and~\eqref{EqEC2} is equal to~\eqref{EqEC3} giving the maximum information sum-rate which can be achieved when the transmitters are using powers $\beta_1 P_1$ and $\beta_2 P_2$ for transmitting information, i.e., $\rho^\star(\beta_1,\beta_2)$ is the information sum-rate optimal correlation coefficient.

\textbf{Existence and Uniqueness of $\rho^\star(\beta_1,\beta_2)$:}
For a fixed power-splitting $(\beta_1,\beta_2) \in (0,1]^2$, let the function $\varphi_{\beta_1,\beta_2}: [0,1]\to \Reals$ denote the difference between the right-hand-side and  the left-hand-side of~\eqref{eq:rhostar}, i.e.,
\begin{IEEEeqnarray}{lcl}
	\varphi_{\beta_1,\beta_2}(\rho)\eqdef&&1+\beta_1\;\SNR_{11}+\beta_2\;\SNR_{12}+2\rho  \sqrt{\beta_1\SNR_{11}\beta_2 \SNR_{12} } \nonumber\\&&- \left(1+\beta_1\;\SNR_{11} (1- \rho^2)\right) \left(1+\beta_2\;\SNR_{12} (1- \rho^2)\right).\IEEEeqnarraynumspace
\end{IEEEeqnarray}
The function $\varphi_{\beta_1,\beta_2}(\rho)$ is continuous in $\rho$ on the closed interval $[0,1]$ and is such that  $\varphi_{\beta_1,\beta_2}(0)<0$ and $\varphi_{\beta_1,\beta_2}(1)>0$, and thus there exists at least one $\rho_0\in (0,1)$ such that $\varphi_{\beta_1,\beta_2}(\rho_0)=0$~\cite[Bolzano's Intermediate Value Theorem (Theorem~5.2.1)]{Calculus}. Furthermore, this solution $\rho_0$ is unique because $\varphi_{\beta_1,\beta_2}(\rho)$ is strictly monotonic on $[0,1]$. This unique solution is $\rho^\star(\beta_1,\beta_2)$.

Note also that the Pearson correlation factor between the NIC signals of both transmitters does not appear in Theorem~\ref{Theorem-EC-Region-MAC-F}. This is mainly because maximum energy transmission occurs using NIC signals that are fully correlated, and thus the corresponding Pearson correlation coefficient is one.  
Similarly, the Pearson correlation factor between the NIC signal of transmitter $i$ and the IC  signal of transmitter $j$, with $j \in \{1,2\}$ and $j\neq i$, does not appear in Theorem~\ref{Theorem-EC-Region-MAC-F} either. This observation stems from the fact that, without loss of optimality, NIC signals can be chosen to be independent of the message indices and  the noise terms.
NIC signals can also be assumed to be known by both the receiver and the transmitters. Hence, the interference they create at the receiver can easily be eliminated using successive decoding. Under this assumption, a power-splitting $(\beta_1,\beta_2) \in [0,1]^2$ guarantees the achievability of non-negative rate pairs $(R_1, R_2)$ satisfying \eqref{EqEC1}-\eqref{EqEC3} by simply using Ozarow's capacity achieving scheme. At the EH, both the IC and NIC signals contribute to the total  harvested energy \eqref{EqEH}. The IC signal  is able to convey at most $\beta_1\SNR_{21} + \beta_2\SNR_{22} + 2 \rho  \sqrt{\beta_1\SNR_{21} \beta_2\SNR_{22}}$ energy-units per channel use, while the NIC signal is able to convey at most $(1-\beta_1)\SNR_{21} + (1-\beta_2)\SNR_{22} + 2 \sqrt{(1-\beta_1)\SNR_{21} (1-\beta_2)\SNR_{22}}$ energy-units per channel use. 
The sum of these two contributions as well as the contribution of the noise at the EH justifies the upper-bound on the energy transmission rate in~\eqref{EqEC4}.

The information-energy capacity region without feedback described by Theorem~\ref{Theorem-EC-Region-MAC-NF} is identical to the information-energy capacity region described by Theorem~\ref{Theorem-EC-Region-MAC-F} in the case in which channel inputs are chosen to be mutually independent, i.e., $\rho = 0$. To prove the achievability of the region presented in Theorem~\ref{Theorem-EC-Region-MAC-NF}, Ozarow's scheme is replaced by the scheme proposed independently by Cover~\cite{COVER75} and Wyner~\cite{WYNER}, in which the channel inputs are independent Gaussian variables.

\subsection{Comments on the Converse}
The proof of the converse to Theorem~\ref{Theorem-EC-Region-MAC-F}  presented in Appendix~\ref{SecProofThm3} is in two steps.
First, it is shown that any information-energy rate triplet $(R_1,R_2,B)\in\set{E}_b^\FB(\SNR_{11},\SNR_{12},\SNR_{21},\SNR_{22})$
must satisfy
\begin{subequations}
	\begin{IEEEeqnarray}{cCl}
		n R_1 &\leqslant& \sum_{t=1}^n I (X_{1,t};Y_{1,t}|X_{2,t})+\epsilon_1^{(n)},\\
		n R_2 &\leqslant& \sum_{t=1}^n I (X_{2,t};Y_{1,t}|X_{1,t})+\epsilon_2^{(n)},\\
		n(R_1+R_2) & \leqslant & \sum_{t=1}^n I(X_{1,t} X_{2,t};Y_{1,t})+\epsilon_{12}^{(n)},\\
		B&\leqslant & \E{B^{(n)}} +\delta^{(n)},\\
		B&\geqslant & b,
	\end{IEEEeqnarray}
\end{subequations}
where $\frac{\epsilon_1^{(n)}}{n},\frac{\epsilon_2^{(n)}}{n}$, $\frac{\epsilon_1^{(n)}}{n}$, and $\delta^{(n)}$ tend to zero as $n$ tends to infinity.
Second, these bounds are evaluated for a general choice of jointly distributed pair of inputs $(X_{1,t}, X_{2,t})$ such that $\E{X_{i,t}}=\mu_{i,t}$, $\Var{X_{i,t}}=\sigma_{i,t}^2$, and  $\textnormal{Cov}[X_{1,t},X_{2,t}]=\lambda_t$, $\forall i \in \{1,2\}$ and $\forall t \in \{1,\dots, n \}$.

The converse to Theorem~\ref{Theorem-EC-Region-MAC-NF} follows the same lines as in the case with feedback, with the assumption that $X_{1,t}$ and $X_{2,t}$ are independent (i.e., $\forall t\in \{1,\dots, n\}$, $\lambda_t=0$).

\subsection{Example}
\begin{figure*}
	\centering
	\hspace*{-2mm}
	\includegraphics[scale=.61]{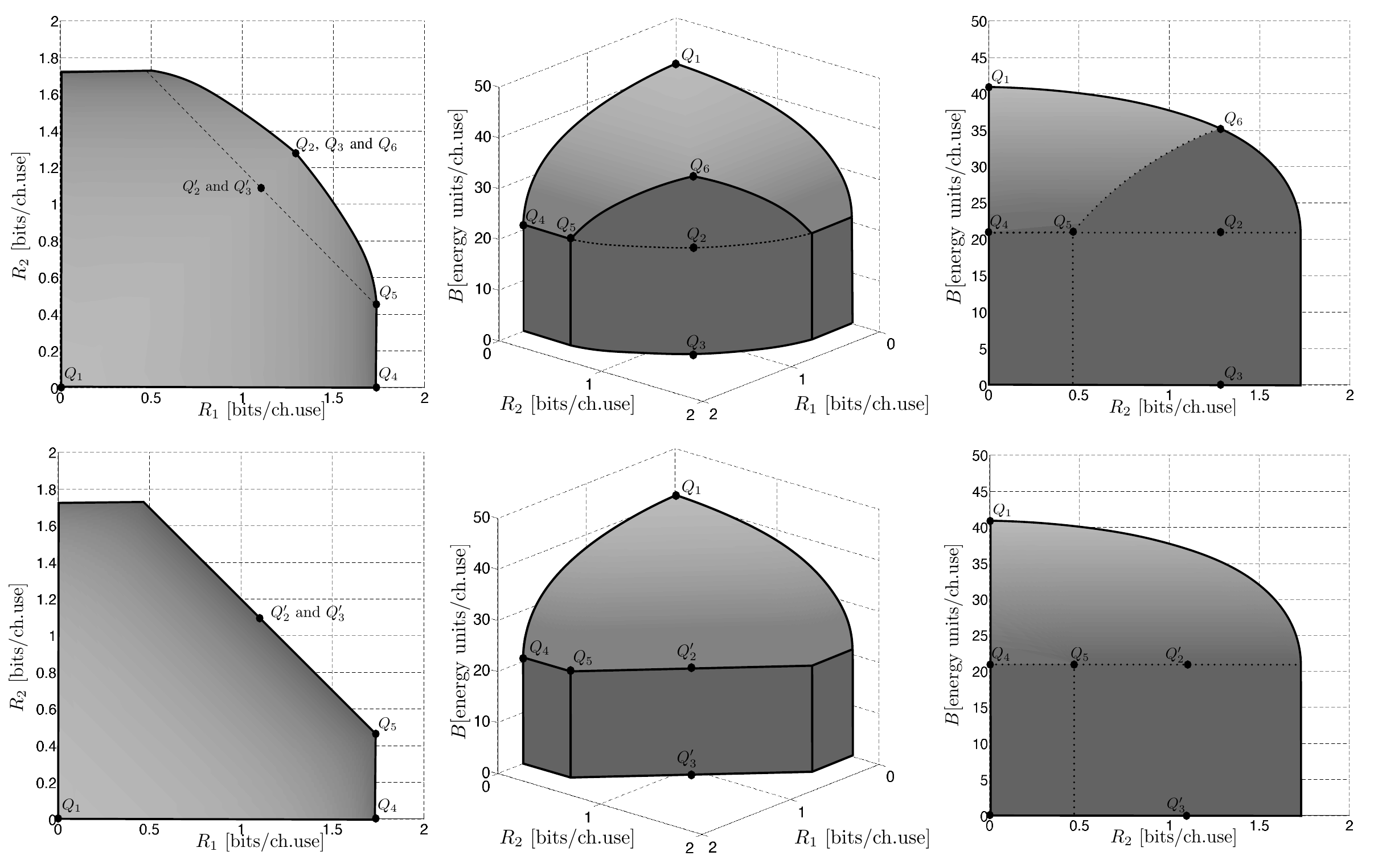}
	\caption{$3$-D representation of the information-energy  capacity  region of the \GMACFB~(top figures) and \GMACB~(bottom figures),    $\set{E}_0^{\mathrm{FB}} \left(10,10,10,10\right)$ and $\set{E}_0 \left(10,10,10,10\right)$, respectively,  with $b=0$, in the coordinate system  $(R_1,R_2,B)$. In each case, the figure in the center is a $3$-D representation of the information-energy capacity region, whereas  left and right figures represent a bi-dimensional view in the $R_1$-$R_2$ and $B$-$R_2$ planes, respectively.  
		Note that $Q_1 = \left(0,0,1 + \SNR_{21} + \SNR_{22} + 2\sqrt{\SNR_{21} \SNR_{22}}\right)$. 
		Points $Q_1$, $Q_2$, $Q_3$, $Q_6$, $Q'_2$, and $Q'_3$ are coplanar and satisfy $R_1 = R_2$.  Points $Q'_2$ and $Q'_3$ satisfy $R_1 = R_2 = \frac{1}{4} \log_2\left(1 + \SNR_{11} + \SNR_{12}  \right)$. 
		Points $Q_2$, $Q_3$, and $Q_6$ are collinear and satisfy $R_1 +  R_2 = \frac{1}{2} \log_2\left(1 + \SNR_{11} + \SNR_{12} + 2\rho^\star(1,1) \sqrt{\SNR_{11} \SNR_{12}} \right)$. 
		The points $Q_2$, $Q_2'$, $Q_4$, and $Q_5$ are coplanar and they satisfy $B = 1 + \SNR_{21} + \SNR_{22}$. In particular, $Q_4= \left(\frac{1}{2}\log_2\left( 1 + \SNR_{11}\right),0,1 + \SNR_{21} + \SNR_{22}\right)$ and $Q_5= \left(\frac{1}{2}\log_2\left( 1 + \SNR_{11}\right),\frac{1}{2}\log_2\left( 1 + \frac{\SNR_{11}}{1 + \SNR_{12}}\right),1 + \SNR_{21} + \SNR_{22}\right)$. 
	}
	\label{Fig3DECregionWithFeedback}
\end{figure*}

Fig.~\ref{Fig3DECregionWithFeedback} shows the information-energy capacity region of the \GMACFB~and the \GMACB, respectively,  with $\SNR_{11}=\SNR_{12}=\SNR_{21}=\SNR_{22} = 10$ and $b=0$.

Therein, in each case, the figure in the center is a {$3$-D} representation of the information-energy capacity region, whereas left and right figures represent a bi-dimensional view in the $R_1$-$R_2$ and $B$-$R_2$ planes, respectively. 
The triplet $Q_1$ with the highest energy transmission rate is $Q_1 = \left(0,0,1 + \SNR_{21} + \SNR_{22} + 2\sqrt{\SNR_{21} \SNR_{22}}\right)$.  The triplets $Q_2$, $Q_2'$, $Q_4$ and $Q_5$ are coplanar and they satisfy $B = 1 + \SNR_{21} + \SNR_{22}$. 
More specifically, $Q_4= \big(\frac{1}{2}\log_2\left( 1 + \SNR_{11}\right),   0,1 + \SNR_{21} + \SNR_{22}\big)$ and $Q_5= \big(\frac{1}{2}\log_2\left( 1 + \SNR_{11}\right), \frac{1}{2}\log_2\left( 1 + \frac{\SNR_{11}}{1 + \SNR_{12}}\right),$ $1 + \SNR_{21} + \SNR_{22}\big)$ are achievable with and without feedback.
In Fig.~\ref{Fig3DECregionWithFeedback}, the triplets $Q_2$, $Q_3$ and $Q_6$ guarantee information transmission at the perfect feedback sum-capacity, i.e., $R_1 +  R_2  = \break\frac{1}{2} \log_2\left(1 + \SNR_{11} + \SNR_{12} + 2\rho^\star(1,1) \sqrt{\SNR_{11} \SNR_{12}} \right)$. 
In the G-MAC(0), the triplets $Q_2$, $Q_3$, and $Q_5$ guarantee information transmission at the sum-capacity without feedback, i.e., $R_1 +  R_2  = \frac{1}{2} \log_2\left(1 + \SNR_{11} + \SNR_{12} \right)$. 

A global comparison of the shape of these two regions is provided in Sec.~\ref{SecMainResultsComments}. This comparison is based on extreme information transmission points, i.e., maximum information  individual and sum rates, given a minimum energy rate. The exact values of these extreme points are derived in Sec.~\ref{SecMainResultsMaxInd} and Sec.~\ref{SecMainResultsMaxSum}.

\section{Maximum Individual Rates Given a Minimum Energy Rate Constraint}
\label{SecMainResultsMaxInd}

In this section, for any fixed non-negative SNRs: $\SNR_{11}$, $\SNR_{12}$, $\SNR_{21}$, and $\SNR_{22}$,  and for any energy rate constraint $b$ at the input of the EH satisfying \eqref{EnergyCst}, the maximum individual information rates of transmitters 1 and 2 in the \GMACFB~and \GMACB~are identified.

Let $\xi: \mathbb{R}_+ \rightarrow [0,1]$ be defined as follows:
\begin{equation}
\label{xi}
\xi(b) \eqdef \frac{\left(b-\left(1+\SNR_{21}+\SNR_{22}\right)\right)^+}{2\sqrt{\SNR_{21}\SNR_{22}}}.
\end{equation}
Note that $\xi(b)$ is the minimum correlation of the channel inputs that is required to achieve the target energy rate $b$. That is, $\xi(b)$ is the solution in $[0,1]$ to 
\begin{equation}
b=1+\SNR_{21}+\SNR_{22}+2 x \sqrt{\SNR_{21}\SNR_{22}}.
\end{equation}

\subsection{Case With Feedback}

The maximum  individual information rate of transmitter $i$, with $i \in \lbrace 1, 2 \rbrace$, denoted by  $R_{i}^\FB(b)$, in the \GMACFB~is the solution to an optimization problem of the form
\begin{IEEEeqnarray}{rCl}
	\label{EqRindiv}
	R_{i}^\FB(b) &=&\;\qquad\qquad\qquad\max_{
		\mathclap{\substack{(R_1,R_2, B)\in\set{E}_b^{\FB} (\SNR_{11},\SNR_{12},\SNR_{21},\SNR_{22})}}} \qquad\qquad\;
	\qquad R_i.
\end{IEEEeqnarray}
The solution to \eqref{EqRindiv} is given by the following proposition. 
\begin{proposition}[Maximum Individual Information Rates of the \GMACFB]%
	\label{PropMaxIndividualRate}
	The maximum individual information rate of transmitter $i$ in a \GMACFB~is given by
	\begin{IEEEeqnarray}{lcl}
		R_i^\FB(b)&=& \frac{1}{2}\log_2\left( 1 +\left(1-\xi(b)^2\right)  \SNR_{1i}\right), \quad i \in \{1,2\},\IEEEeqnarraynumspace
	\end{IEEEeqnarray}
	with  $\xi(b) \in [0,1]$ defined in \eqref{xi}.
\end{proposition}

\begin{IEEEproof}
	The proof of Proposition~\ref{PropMaxIndividualRate} is provided in Appendix~\ref{ProofPropMaxIndividualRate}.
\end{IEEEproof}

\subsection{Case Without Feedback}

The maximum  individual information rate of transmitter $i$ in the \GMACB, with $i \in \lbrace 1, 2 \rbrace$, denoted by $R_i^\NF(b)$, is the solution to an optimization problem of the form
\begin{IEEEeqnarray}{rCl}
	\label{EqRindivNF}
	R_{i}^\NF(b)
	&=&\qquad\qquad\qquad\max_{
		\mathclap{\substack{(R_1,R_2, B)\in\set{E}_b (\SNR_{11},\SNR_{12},\SNR_{21},\SNR_{22})}}} \qquad\qquad
	\qquad R_i.
\end{IEEEeqnarray}

The solution to \eqref{EqRindivNF} is given by the following proposition. 
\begin{proposition}[Maximum Individual Information Rates of the \GMACB]
	\label{PropMaxIndividualRateNF}
	The maximum individual information rate of transmitter $i$ in a \GMACB~is given by
	\begin{IEEEeqnarray}{lcl}
		R_i^\NF(b)&=& R_i^\FB(b),\quad i \in \{1,2\}.
	\end{IEEEeqnarray}
\end{proposition}
\begin{IEEEproof}
	The proof of Proposition~\ref{PropMaxIndividualRateNF} is presented in Appendix~\ref{ProofPropMaxIndividualRateNF}.	\end{IEEEproof}
That is, the maximum individual information rates in the \GMACFB~and in the \GMACB~coincide.

\section{Maximum Information Sum-Rate Given a Minimum Energy Rate Constraint}
\label{SecMainResultsMaxSum}
In this section, for any fixed non-negative $\SNR_{11}$, $\SNR_{12}$, $\SNR_{21}$, and $\SNR_{22}$,  and for any $b$ satisfying \eqref{EnergyCst}, the information sum-capacity (i.e., the maximum information sum-rate) is identified in the \GMACFB~and in the \GMACB.

\subsection{Case With Feedback}	

The perfect feedback information sum-capacity  $R_{\mathrm{sum}}^\FB(b)$ of the {\GMACFB}~is the solution to an optimization problem of the form
\begin{IEEEeqnarray}{rCl}
	\label{EqRsum}
	R_{\mathrm{sum}}^\FB( 
	b)
	&=&\qquad\qquad\qquad\max_{
		\mathclap{\substack{(R_1,R_2, B)\in\set{E}_b^{\mathrm{FB}} (\SNR_{11},\SNR_{12},\SNR_{21},\SNR_{22})}}} \qquad\qquad
	\qquad R_1 + R_2.\IEEEeqnarraynumspace
\end{IEEEeqnarray}
The solution to \eqref{EqRsum} is given by the following proposition. 
\begin{proposition}[Information Sum-Capacity of the \GMACFB]
	\label{PropMaxSumRate}
	The information sum-capacity of the\break\GMACFB~is
	\begin{enumerate}
		\item $\forall b \in {\left[\hspace*{-0.2mm}0,\hspace*{-0.5mm}1\hspace*{-0.5mm}+\hspace*{-0.5mm}\SNR_{21}\hspace*{-0.5mm}+\hspace*{-0.5mm}\SNR_{22}\hspace*{-0.5mm}+\hspace*{-0.5mm}2\rho^\star(1\hspace*{-0.5mm},\hspace*{-0.5mm}1) \hspace*{-0.5mm}\sqrt{\SNR_{21}\SNR_{22}} \right]}$, 
		\begin{IEEEeqnarray}{l}
			\hspace*{-4mm}R_{\mathrm{sum}}^\FB(b) =
			\hspace*{-4mm}
			\frac{1}{2} \log_2(1\hspace*{-1mm} + \hspace*{-1mm} \SNR_{11}\hspace*{-1mm} +\hspace*{-1mm} \SNR_{12} \hspace*{-1mm}+\hspace*{-1mm} 2 \rho^\star(1,1) \sqrt{\SNR_{11}\SNR_{12}});\hspace*{-1mm}\label{eq:28}\IEEEeqnarraynumspace 
		\end{IEEEeqnarray}
		\item $\forall b \in \big(1 + \SNR_{21} + \SNR_{22}+ 2 \rho^\star(1,1) \sqrt{\SNR_{21} \SNR_{22}}, 1 +  \SNR_{21} + \SNR_{22}+ 2 \sqrt{\SNR_{21} \SNR_{22}} \big)$,
		\begin{IEEEeqnarray}{lcl}
			\label{eq:29}
			R_{\mathrm{sum}}^\FB(b) &=&
			\frac{1}{2} \log_2(1 +  (1-\xi(b)^2) \SNR_{11})+ \frac{1}{2}\log_2(1 + (1-\xi(b)^2) \SNR_{12}); \quad
		\end{IEEEeqnarray}
		\item  $\forall b \in \big[1 +  \SNR_{21} + \SNR_{22} + 2 \sqrt{\SNR_{21} \SNR_{22}}, \infty]$, 
		\begin{IEEEeqnarray}{l}
			R_{\mathrm{sum}}^\FB(b) =  0,
		\end{IEEEeqnarray}	
	\end{enumerate}
	where  $\rho^\star(1,1)$ denotes the unique solution in $(0,1)$ to \eqref{eq:rhostar} with $\beta_1=\beta_2=1$ and the function $\xi(b)$ is defined in \eqref{xi}.
\end{proposition}
\begin{IEEEproof}
	The proof of Proposition~\ref{PropMaxSumRate} is presented in Appendix~\ref{SecProofThm4}.\end{IEEEproof}

\subsection{Case Without Feedback}

The information sum-capacity  $R_{\mathrm{sum}}^\NF(b)$  of the \GMACB~is the solution to an optimization problem of the form
\begin{IEEEeqnarray}{rCl}
	\label{EqRsumNF}
	R_{\mathrm{sum}}^\NF(b) 
	&=&\qquad\qquad\qquad\max_{
		\mathclap{\substack{(R_1,R_2, B)\in\set{E}_b (\SNR_{11},\SNR_{12},\SNR_{21},\SNR_{22})}}} 
	\qquad\qquad\qquad R_1 + R_2.\IEEEeqnarraynumspace
\end{IEEEeqnarray}
The solution to~\eqref{EqRsumNF} is given by the following proposition.

\begin{proposition}[Information Sum-Capacity of the \GMACB]
	\label{PropMaxSumRateNF}
	The information sum-capacity of the \GMACB~is%
	\begin{enumerate}
		\item $\forall b\in \Big[0, 1+\SNR_{21}+\SNR_{22}+2 \sqrt{\SNR_{21}\SNR_{22}} \min\left\{ \sqrt{\frac{\SNR_{12}}{\SNR_{11}}},\sqrt{\frac{\SNR_{11}}{\SNR_{12}}}\right\} \Big]$
		\begin{IEEEeqnarray}{lCl}
			R_{\mathrm{sum}}^\NF(b)&=&\frac12 \hspace*{-.5mm}\log_2\hspace*{-1mm}\left(\hspace*{-.5mm}1\hspace*{-.5mm}+\hspace*{-.5mm}\SNR_{11}\hspace*{-.5mm}+\hspace*{-.5mm}\SNR_{12}\hspace*{-1mm}-\hspace*{-1mm} 2 \xi(b)  \hspace*{-.5mm}\sqrt{\hspace*{-.5mm}\SNR_{11}\SNR_{12}}\right)\hspace*{-1mm},
		\end{IEEEeqnarray}
		\item $\forall b\in \Big(1+\SNR_{21}+\SNR_{22}+2 \sqrt{\SNR_{21}\SNR_{22}} \min\left\{ \sqrt{\frac{\SNR_{12}}{\SNR_{11}}},\sqrt{\frac{\SNR_{11}}{\SNR_{12}}}\right\},1+\SNR_{21}+\SNR_{22}+2 \sqrt{\SNR_{21}\SNR_{22}}\Big]$
		\begin{IEEEeqnarray}{lCl}
			\label{eq:35}
			R_{\mathrm{sum}}^\NF(b) &=&\frac12 \log_2\left(1+\left(1-\xi(b)^2\right)\SNR_{1i}\right),
		\end{IEEEeqnarray}
		with $\ds i=\argmax_{k\in \{1,2\}} \;\SNR_{1k}$, \item $\forall b\in \left[1+\SNR_{21}+\SNR_{22}+2 \sqrt{\SNR_{21}\SNR_{22}},\infty\right]$
		\begin{IEEEeqnarray}{lCl}
			R_{\mathrm{sum}}^\NF(b) &=&0,
		\end{IEEEeqnarray}
	\end{enumerate}
	with the function $\xi(b)$ defined in \eqref{xi}.

\end{proposition}

\begin{IEEEproof}
	The proof is presented in Appendix~\ref{SecProofProp2}.
\end{IEEEproof}		

From Propositions~\ref{PropMaxSumRate} and \ref{PropMaxSumRateNF}, it can be seen that in the case with feedback, both users might transmit information and energy simultaneously as feedback creates signal correlation, which allows the system to meet the minimum energy rate. That is, the correlation induced by the use of the feedback is beneficial to both information transmission and energy transmission. 
Alternatively, in the case without feedback, artificial correlation via common randomness is required to meet the energy rate constraint. Such a correlation only benefits the energy transmission task and comes at the expense of the information transmission task as the information sum-rate is necessarily reduced. For instance, one way of achieving \eqref{eq:35} is when the transmitter with the lowest SNR uses common randomness at its maximum power (transmits only energy), while the other transmitter transmits both energy and information.

\begin{remark}
	\label{rem:TS}
	Optimally alternating transmission of energy and information does not always	%
	achieve information sum-capacity of the \GMACB~for a given minimum received energy rate constraint $b$. 
\end{remark}	
To verify Remark~\ref{rem:TS}, consider the sum-rate optimization problem proposed in~\cite{Fouladgar-CL-2012} in which both users alternate between information and energy transmission. Specifically, during a fraction of time $\lambda\in [0,1]$, transmitter $i$ sends an IC signal with  power $P'_i$ and during the remaining fraction of time it sends an NIC signal with power $P''_i$. Thus, the sum-rate optimal time-sharing parameter $\lambda$ and power control vector $(P'_1,P'_2,P''_1,P''_2)$ are solutions to the optimization problem
\begin{subequations}
	\label{eq:timesharing}
	\begin{IEEEeqnarray}{l}
		\max_{(\lambda,P'_1,P''_1,P'_2,P''_2) \in [0,1]\times \Reals_+^4} \frac{\lambda}{2}\log_2\left(1+h_{11}^2 P'_1+h_{12}^2 P'_2\right)\IEEEeqnarraynumspace\\
		\textnormal{subject to}:\nonumber\\ 
		\lambda P'_i + (1 - \lambda) P''_i\leqslant  P_i,\quad i \in \{1,2\}\label{eq:TSconstraints}\\
		1\hspace*{-1mm}+\hspace*{-1mm} \lambda(h_{21}^2 P'_1\hspace*{-1mm}+\hspace*{-1mm}h_{22}^2 P'_2)
		\hspace*{-1mm}+\hspace*{-1mm}(1-\lambda)
		(h_{21}\hspace*{-.5mm}\sqrt{P''_1}\hspace*{-1mm}+\hspace*{-1mm}h_{22}\hspace*{-.5mm} \sqrt{P''_2})^2\hspace*{-1mm}\geqslant\hspace*{-.5mm} b, 
	\end{IEEEeqnarray}		
\end{subequations}
where $P_i$ is the total power budget of transmitter $i$.

For any feasible choice of $(\lambda, P'_1,P''_1, P'_2, P''_2)$, by the concavity of the logarithm, it follows that:
\begin{equation}\label{eq:cmp}
\hspace*{-2mm}\frac{\lambda}{2}\hspace*{-.5mm}\log_2\hspace*{-.5mm}(\hspace*{-.5mm}1\hspace*{-.5mm}+\hspace*{-.5mm}h_{11}^2 \hspace*{-.5mm}P'_1\hspace*{-.5mm}+\hspace*{-.5mm}h_{12}^2 P'_2) 
\hspace*{-.5mm}\leqslant\hspace*{-.5mm}\frac{1}{2}\hspace*{-.5mm}\log_2\hspace*{-.5mm}\left(\hspace*{-.5mm}1\hspace*{-.5mm}+\hspace*{-.5mm}\lambda\hspace*{-.5mm} \left(h_{11}^2 P'_1\hspace*{-.5mm}+\hspace*{-.5mm}h_{12}^2 P'_2\right)\right)\hspace*{-.5mm}.
\end{equation}	 
Note that for $\lambda \neq 1$, the inequality in~\eqref{eq:cmp} is strict and the rate $\frac{1}{2}\log_2\left(1+\lambda \left(h_{11}^2 P'_1+h_{12}^2 P'_2\right)\right)$ is always achievable by a power-splitting scheme in which $\beta_i=\lambda \frac{P'_i}{P_i}$, with $i \in \{1,2\}$, for any optimal tuple $(\lambda, P'_1,P''_1, P'_2,P''_2)$ in  \eqref{eq:timesharing}. This shows that the maximum information sum-rate achieved via alternating energy and information transmission is always bounded away from the information sum-capacity (Proposition \ref{PropMaxSumRateNF}). When $\lambda = 1$, exclusively transmitting information  satisfies  the energy rate constraint, i.e., $b \in [0, 1 + \SNR_{21} + \SNR_{22}]$.

\section{Comments on the Shape of the Information-Energy Capacity Region}
\label{SecMainResultsComments}

In this section, observations on the shape of the volumes  $\set{E}_0^\FB \left(\SNR_{11},\SNR_{12},\SNR_{21},\SNR_{22} \right)$ and\break
$\set{E}_0\left(\SNR_{11},\SNR_{12},\SNR_{21},\SNR_{22}\right)$ are presented.

For a given $k\in \mathbb{N}$, let $\mathcal{B}(b_k) \subset \mathbb{R}^{2}_{+}$ be a two-dimensional set of the form
\begin{IEEEeqnarray}{lcl} 
	\mathcal{B}(b_k)  = \Big\lbrace   (R_1, R_2) \in \mathbb{R}^{2}_{+}:  R_i \leqslant \frac{1}{2} \log_2 \left( 1 + \left(1-\xi(b_k)^2\right)\SNR_{1i} \right), i \in \lbrace 1, 2 \rbrace \Big\rbrace.
\end{IEEEeqnarray}

\subsection{Case With Feedback}
\begin{figure*}
	\centering
	\includegraphics[scale=.5]{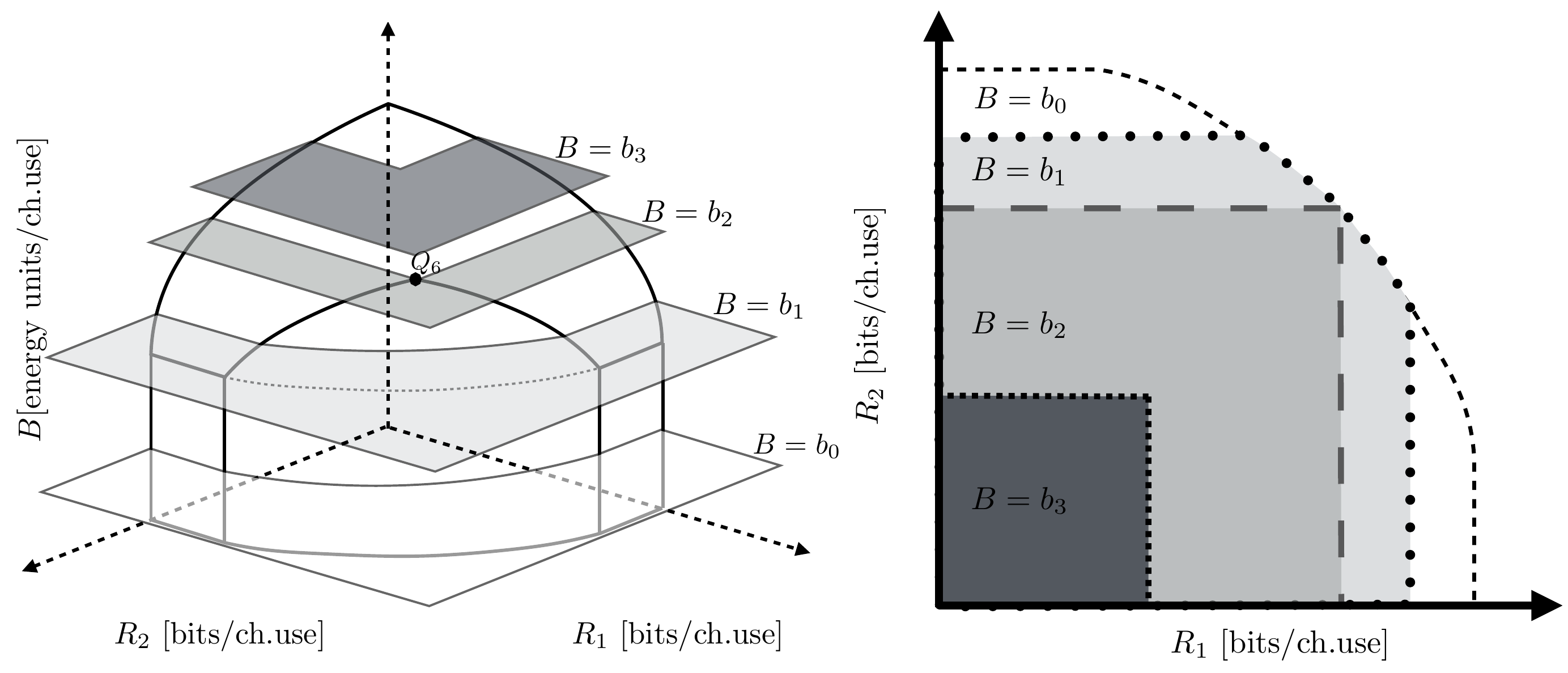}
	\caption{Intersection of the the information-energy capacity region of the \GMACF$(O)$, $\set{E}_0^\FB \left(\SNR_{11},\SNR_{12},\SNR_{21},\SNR_{22} \right)$, with the planes $B=b_0$, $B=b_1$, $B=b_2$ and $B=b_3$ where 
		$b_0 \in \left[0, 1 + \SNR_{21} + \SNR_{22} \right]$, $b_1 \in \left[1 + \SNR_{21} + \SNR_{22}, 1 + \SNR_{21} + \SNR_{22} + 2\rho^\star(1 , 1)  \sqrt{\SNR_{21}\SNR_{22}} \right]$, $b_2 = 1 + \SNR_{21} + \SNR_{22} + 2\rho^\star(1,1)  \sqrt{\SNR_{21}\SNR_{22}} $, and $b_3 \in \left[1+ \SNR_{21} + \SNR_{22} + 2\rho^\star(1,1)  \sqrt{\SNR_{21}\SNR_{22}},  1+ \SNR_{21} + \SNR_{22} + 2  \sqrt{\SNR_{21}\SNR_{22}}\right]$.}
	\label{FigCuts}
\end{figure*}

Fig.~\ref{FigCuts} shows a general example of the intersection of the volume\break$\set{E}_0^{\FB} \left(\SNR_{11},\SNR_{12},\SNR_{21},\SNR_{22}\right)$, in the Cartesian coordinates $(R_1,R_2,B)$, with the planes $B =b_k$, with $k\in \{0,1,2,3\}$, such that $b_0 \in \big[0, 1 + \SNR_{21} + \SNR_{22} \big]$, $b_1 \in \big[1 + \SNR_{21} + \SNR_{22}, 1 + \SNR_{21} + \SNR_{22} + 2\rho^\star(1 , 1)  \sqrt{\SNR_{21}\SNR_{22}} \big]$, $b_2 = 1 + \SNR_{21} + \SNR_{22} + 2\rho^\star(1,1)  \sqrt{\SNR_{21}\SNR_{22}} $, and $b_3 \in \big[1+ \SNR_{21} + \SNR_{22} + 2\rho^\star(1,1)  \sqrt{\SNR_{21}\SNR_{22}},  1+ \SNR_{21} + \SNR_{22} + 2  \sqrt{\SNR_{21}\SNR_{22}}\big]$.

\textbf{Case 1: $\mathbf{b_0 \in \left[0,1+\SNR_{21}+\SNR_{22}\right]}$.} 
In this case, any intersection of the volume\break$\set{E}_0^\FB \left(\SNR_{11},\SNR_{12},\SNR_{21},\SNR_{22}\right)$, in the Cartesian coordinates $(R_1,R_2,B)$, with a plane $B = b_0$ corresponds to the set of triplets $(R_1,R_2, b_0)$, in which the corresponding pairs $(R_1,R_2)$ form a set that is identical to the information capacity region of the \GMACF~(without EH), denoted by $\set{C}_\FB(\SNR_{11},\SNR_{12})$. Note that this intersection is the base of the information-energy capacity region $\set{E}_{b_0}^\FB\hspace*{-1mm} \left(\SNR_{11},\SNR_{12},\SNR_{21},\SNR_{22}\right)$ of the \GMACF$(b_0)$. In this case, $\xi(b_0) = 0$, and thus from Proposition~\ref{PropMaxIndividualRate} and Proposition~\ref{PropMaxSumRate}, the energy constraint does not add any additional bound on the individual rates and sum-rate other than \eqref{EqEC1}, \eqref{EqEC2}, and \eqref{EqEC3}. That is, the minimum energy transmission rate requirement can always be met by exclusively transmitting information. 

\textbf{Case 2: $\mathbf{b_1 \in \big(1+\SNR_{21}+\SNR_{22}, 1+\SNR_{21}+\SNR_{22}}$ $\mathbf{+ 2\rho^{\star}(1,1)\sqrt{\SNR_{21} \SNR_{22}}\big]}$.} 
In this case, any intersection of the volume $\set{E}_0^\FB \left(\SNR_{11},\SNR_{12},\SNR_{21},\SNR_{22}\right)$  with a plane $B=b_1$ is a set of triplets $(R_1,R_2, b_1)$ for which the corresponding pairs $(R_1,R_2)$ satisfy $(R_1,R_2) \in \mathcal{B}(b_1) \cap \set{C}_\FB(\SNR_{11},\SNR_{12})$,
which forms a strict subset of  $\set{C}_\FB(\SNR_{11},\SNR_{12})$. This intersection coincides with the base of the information-energy capacity region  $\set{E}_{b_1}^\FB \left(\SNR_{11},\SNR_{12},\SNR_{21},\SNR_{22}\right)$  of the \GMACF$(b_1)$. Note that
$\xi(b_1)>0$, and thus from Proposition~\ref{PropMaxIndividualRate}, the energy constraint limits the individual rates. That is, transmitter $i$'s individual information rate is bounded away from $\frac{1}{2}\log_2\left( 1 + \SNR_{1i} \right)$. Nevertheless, it is important to highlight that in this case, $\xi(b_1)\leqslant  \rho^{\star}(1,1)$, and thus the individual rates $R_1=\frac12 \log_2\left(1+ \left(1- \left( \rho^{\star}(1,1) \right)^2\right) \SNR_{11}\right)$ and $R_2=\frac12 \log_2\left(1+ \left(1- \left( \rho^{\star}(1,1) \right)^2\right) \SNR_{12}\right)$ are always achievable. Hence,  this intersection always includes the triplet $(R_1,R_2, b_1)$, with $R_1\hspace*{-.5mm} + \hspace*{-.5mm}R_2 \hspace*{-1mm}=\hspace*{-1mm}\frac{1}{2}\hspace*{-.5mm} \log_2\hspace*{-1mm}\left(\hspace*{-.5mm}1\hspace*{-.5mm}+\hspace*{-.5mm}\SNR_{11}\hspace*{-.5mm}+\hspace*{-.5mm}\SNR_{12}\hspace*{-.5mm} +\hspace*{-.5mm} 2\rho^{\star}(1,1)\sqrt{\SNR_{11} \SNR_{12}}\right)\break=R_{\mathrm{sum}}^\FB(b_1)=R_{\mathrm{sum}}^\FB(0)$.
That is, the power-split $\beta_1 = \beta_2 = 1$ is always feasible. Note that the intersection of the volume $\set{E}_0^\FB \left(\SNR_{11},\SNR_{12},\SNR_{21},\SNR_{22}\right)$
with the plane $B=b_2$ is a particular case of this regime.

\textbf{Case 3: $\mathbf{b_3 \in \big(1+\SNR_{21}+\SNR_{22} + 2\rho^{\star}(1,1)\sqrt{\SNR_{21} \SNR_{22}},1+\SNR_{21}+\SNR_{22}+ }$\break$\mathbf{2 \sqrt{\SNR_{21} \SNR_{22}}\big]}$.} 
In this case, any intersection of the volume $\set{E}_0^{\FB}\hspace*{-1mm} \left(\hspace*{-.5mm}\SNR_{11},\SNR_{12},\SNR_{21},\SNR_{22}\right)$  with a plane $B = b_3$ is a set of triplets $(R_1,R_2, b_3)$ for which the corresponding pairs $(R_1,R_2)$ satisfy $(R_1,R_2) \in \set{B}(b_3)= \mathcal{B}(b_3) \cap \set{C}_{\mathrm{FB}}(\SNR_{11},\SNR_{12})$, which is a strict subset of $\set{C}_{\mathrm{FB}}(\SNR_{11},\SNR_{12})$. This intersection coincides with the base of the information-energy capacity region\break$\set{E}_{b_3}^\FB \left(\SNR_{11},\SNR_{12},\SNR_{21},\SNR_{22}\right)$  of the \GMACF$(b_3)$. Note that $\rho^{\star}(1,1) < \xi(b_3) \leqslant 1$, and thus from Proposition~\ref{PropMaxIndividualRate}, the individual information rates are limited by\break${R_i\leqslant \frac12\log_2\left( 1 + \left( 1- \xi(b_3)^2 \right) \SNR_{1i} \right)}<\frac12\log_2\left( 1 + \left( 1- \left( \rho^{\star}(1,1) \right)^2 \right) \SNR_{1i} \right)$.  
For any $b_3 > 1+\SNR_{21}+\SNR_{22} + 2\rho^{\star}(1,1)\sqrt{\SNR_{21} \SNR_{22}}$, the set $\mathcal{B}(b_3)$ monotonically shrinks with $b_3$.
Consequently, for these values of $b_3$, there exists a loss of sum-rate and $R_{\mathrm{sum}}^\FB(0)$ is not achievable. 
Nonetheless, note that $R_{\mathrm{sum}}^\FB(b_3)$ is a  continuous function in $b_3$. When $b_3=1 + \SNR_{21} + \SNR_{22}+ 2{ (\rho^\star(1,1)+\epsilon)} \sqrt{\SNR_{21} \SNR_{22}},$ for some $\epsilon >0$, it holds that $\xi(b_3)=\rho^\star(1,1)+\epsilon$. Substituting this into \eqref{eq:29} and  
taking the limit when $\epsilon$ tends to $0$, by the definition of $\rho^\star(1,1)$, the resulting value is given by \eqref{eq:28}. Clearly, the maximum energy rate is achieved when $\beta_1 = \beta_2 = 0$, which implies that no information is conveyed from the transmitters to the receiver.

\subsection{Case Without Feedback}
\begin{figure*}[ht]
	\centering
	\includegraphics[scale=.5]{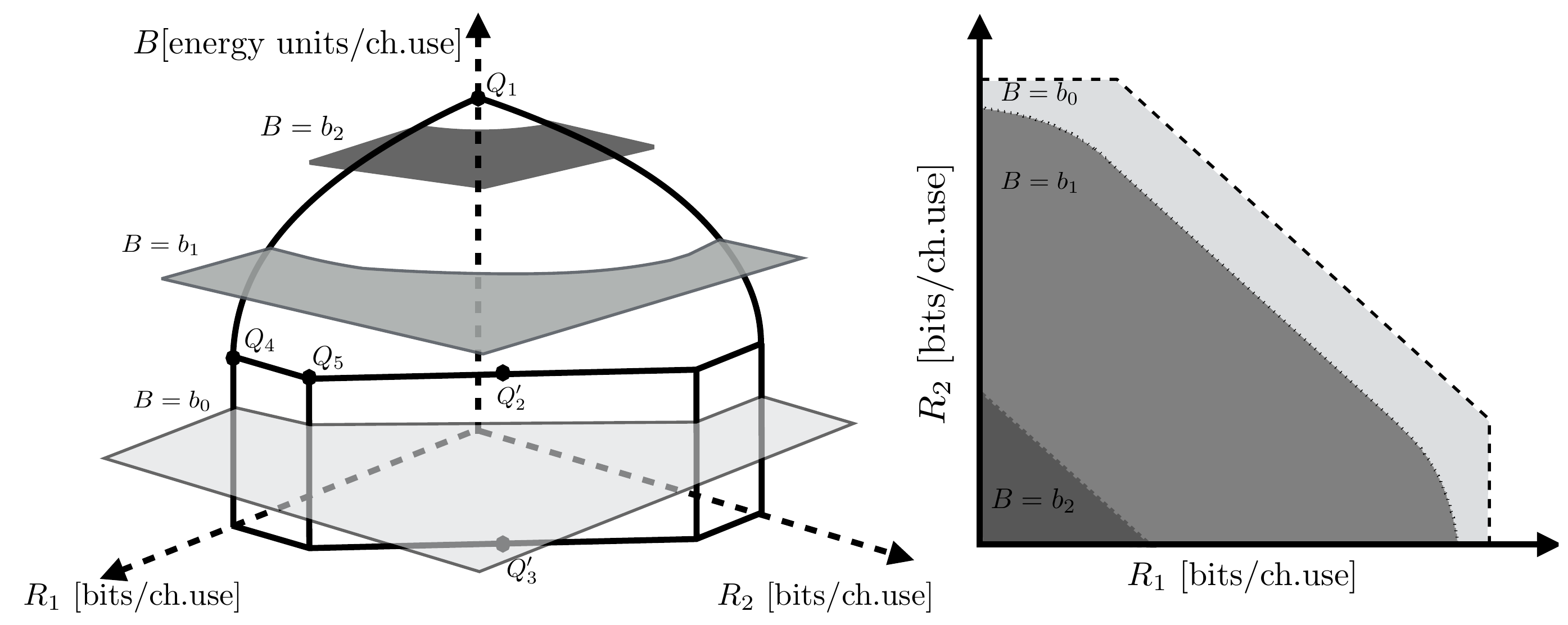}
	\caption{Intersection of the information-energy capacity region of the \GMAC$(0)$, $\set{E}_0\left(\SNR_{11},\SNR_{12},\SNR_{21},\SNR_{22} \right)$, with the planes $B=b_0$, $B=b_1$, and $B=b_2$, where $b_0 \in \left[0,1+\SNR_{21}+\SNR_{22}\right] $, $b_1 \in \left(1+\SNR_{21}+\SNR_{22},1+\SNR_{21}+\SNR_{22} + 2\sqrt{\SNR_{21} \SNR_{22}}  \min\left\{ \sqrt{\frac{\SNR_{12}}{\SNR_{11}}},\sqrt{\frac{\SNR_{11}}{\SNR_{12}}}\right\}     \right]$, and $b_2 \in \Bigr(1+\SNR_{21}+\SNR_{22} + 2\sqrt{\SNR_{21} \SNR_{22}}  \min\left\{ \sqrt{\frac{\SNR_{12}}{\SNR_{11}}},\sqrt{\frac{\SNR_{11}}{\SNR_{12}}}\right\}    ,1+\SNR_{21}+\SNR_{22} +2\sqrt{\SNR_{21} \SNR_{22}}\Bigr]$.
		\label{FigCutsNF}}
\end{figure*}

Fig.~\ref{FigCutsNF} shows a general example of  the intersection of the volume\break$\set{E}_0 \left(\SNR_{11},\SNR_{12},\SNR_{21},\SNR_{22}\right)$, in the Cartesian coordinates $(R_1,R_2,B)$, with the planes $B =b_k$, with $k\in \{0,1,2\}$, such that $b_0 \in [0,1+\SNR_{21}+\SNR_{22}] $, $b_1 \in \Big(1+\SNR_{21}+\SNR_{22},1+\SNR_{21}+\SNR_{22} + 2\sqrt{\SNR_{21} \SNR_{22}}  \min\left\{ \sqrt{\frac{\SNR_{12}}{\SNR_{11}}},\sqrt{\frac{\SNR_{11}}{\SNR_{12}}}\right\}\Big]$, and $b_2 \in \Big(1+\SNR_{21}+\SNR_{22}+ 2\sqrt{\SNR_{21} \SNR_{22}}  \min\left\{ \sqrt{\frac{\SNR_{12}}{\SNR_{11}}},\sqrt{\frac{\SNR_{11}}{\SNR_{12}}}\right\}    ,1+\SNR_{21}+\SNR_{22} +2\sqrt{\SNR_{21} \SNR_{22}}\Big]$.

\textbf{Case 1: $\mathbf{b_0 \in [0,1+\SNR_{21}+\SNR_{22}]}$.} In this case, any intersection of the volume\break$\set{E}_0\left(\SNR_{11},\SNR_{12},\SNR_{21},\SNR_{22}\right)$, in the Cartesian coordinates $(R_1,R_2,B)$, with a plane $B = b_0$ corresponds to the set of triplets $(R_1,R_2, b_0)$, in which the corresponding pairs $(R_1,R_2)$ form a set that is identical to the information capacity region of the \GMAC~(without EH), denoted by $\set{C}(\SNR_{11},\SNR_{12})$. 
This intersection is the base of the information-energy capacity region $\set{E}_{b_0} \left(\SNR_{11},\SNR_{12},\SNR_{21},\SNR_{22}\right)$ of the \GMAC$(b_0)$. Note that $\xi(b_0)=0$, and thus from Proposition~\ref{PropMaxIndividualRateNF} and Proposition~\ref{PropMaxSumRateNF},  it holds that $R_{i}^\NF(b_0) =\frac12 \log_2\left(1+\SNR_{1i}\right)$, for $i\in \{1,2\}$,  and $R_{\mathrm{sum}}^\NF(b_0) =\frac12 \log_2\left(1+\SNR_{11}+\SNR_{12}\right)$.
Hence, exclusively transmitting information  is enough for satisfying the energy rate constraint~$b_0$.

\textbf{Case 2: $\mathbf{b_1\hspace*{-.8mm} \in  \hspace*{-.8mm}\Big(\hspace*{-.8mm}1\hspace*{-.8mm}+\hspace*{-.8mm}\SNR_{21}\hspace*{-.8mm}+\hspace*{-.8mm}\SNR_{22}\hspace*{-.8mm},\hspace*{-.8mm}1\hspace*{-.8mm}+\hspace*{-.8mm}\SNR_{21}\hspace*{-.8mm}+\hspace*{-.8mm}\SNR_{22}\hspace*{-.8mm} + \hspace*{-.8mm}2\hspace*{-.8mm}\sqrt{\hspace*{-.8mm}\SNR_{21}\hspace*{-.5mm} \SNR_{22}}\hspace*{-.8mm}  \min\hspace*{-.8mm}\left\{\hspace*{-.8mm} \sqrt{\hspace*{-.8mm}\frac{\SNR_{12}}{\SNR_{11}}},\hspace*{-.8mm}\sqrt{\hspace*{-.8mm}\frac{\SNR_{11}}{\SNR_{12}}}\right\}\hspace*{-.8mm} \Big]}\hspace*{-.5mm}$.}
In this case, any intersection of the volume $\set{E}_0\left(\SNR_{11},\SNR_{12},\SNR_{21},\SNR_{22}\right)$, in the Cartesian coordinates $(R_1,R_2,B)$, with a plane $B = b_1$ corresponds to the set of triplets $(R_1,R_2, b_1)$ in which the corresponding pairs $(R_1,R_2)$ form a set that is equivalent to a strict subset of the information capacity region of the \GMAC~$\set{C}(\SNR_{11},\SNR_{12})$. This intersection is the base of the information-energy capacity region $\set{E}_{b_1} \left(\SNR_{11},\SNR_{12},\SNR_{21},\SNR_{22}\right)$ of the \GMAC$(b_1)$. Note that $\xi(b_1)>0$, and thus from Proposition~\ref{PropMaxIndividualRateNF} and Proposition~\ref{PropMaxSumRateNF}, $R_{i}^\NF(b_1)$  and $R_{\mathrm{sum}}^\NF(b_1)$ decrease with $b_1$. This is mainly due to the fact that part of each transmitter's power budget is dedicated to the transmission of energy.   
Furthermore,  the information sum-rate optimal strategy involves information transmission at both users since the sum-capacity is strictly larger than the maximum individual rate of the user with the highest SNR. 

\textbf{Case~3: $\mathbf{b_2\hspace*{-.5mm}\in\hspace*{-.5mm} \Big(\hspace*{-.5mm}1+\SNR_{21}+\SNR_{22} +}$ $\mathbf{ 2\sqrt{\SNR_{21} \SNR_{22}}\min\hspace*{-1mm}\left\{\hspace*{-1mm} \sqrt{\frac{\SNR_{12}}{\SNR_{11}}},\hspace*{-.6mm}\sqrt{\frac{\SNR_{11}}{\SNR_{12}}}\right\}\hspace*{-.6mm} , 1\hspace*{-1mm}+\hspace*{-1mm}\SNR_{21}}$\break $\mathbf{+\SNR_{22}+2\sqrt{\SNR_{21} \SNR_{22}}\Big]}$.}  
In this case,  
any intersection of the volume\break$\set{E}_0\left(\SNR_{11},\SNR_{12},\SNR_{21},\SNR_{22}\right)$, in the Cartesian coordinates $(R_1,R_2,B)$, with a plane $B = b_2$ corresponds to the set of triplets $(R_1,R_2, b_2)$ in which the corresponding pairs $(R_1,R_2)$ form a set that is equivalent to a strict subset of the information capacity region of the \GMAC,~$\set{C}(\SNR_{11},\SNR_{12})$. This intersection is the base of the information-energy capacity\break$\set{E}_{b_2} \left(\SNR_{11},\SNR_{12},\SNR_{21},\SNR_{22}\right)$ region of the \GMAC$(b_2)$. The information sum-capacity corresponds to the maximum individual rate (Proposition~\ref{PropMaxIndividualRateNF}) of the transmitter with the highest SNR. That is, in order to maximize the information sum-rate, it is optimal to have information transmission exclusively at the stronger user with the highest SNR. The transmitter with the weakest SNR uses all its power budget to exclusively transmit energy. Note that when the receiver and the EH are co-located and when the channel is symmetric, this is not observed.

\section{Energy Transmission Enhancement With Feedback}
\label{SecMainResultsEnergy}
\begin{figure}[t]
	\centering
	\includegraphics[scale=0.9]{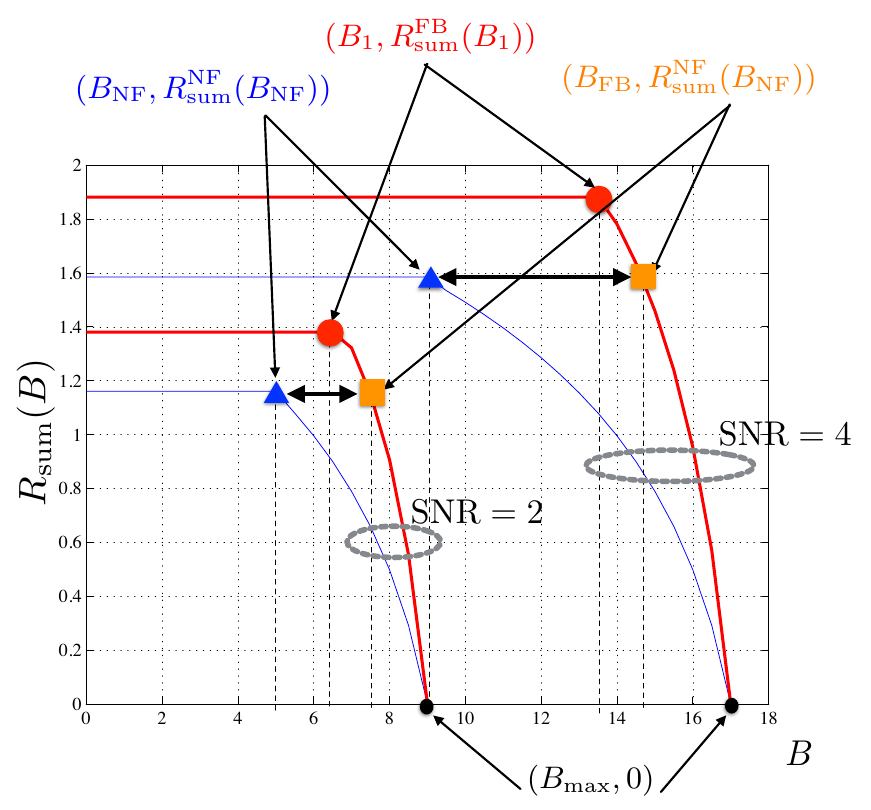}
	\caption{Information sum-capacity of the symmetric two-user memoryless \GMACF$(0)$  (thick red line) and \GMAC$(0)$ (thin blue line),  with co-located receiver and EH, with  $\SNR_{11}=\SNR_{12}=\SNR_{21}=\SNR_{22}=\SNR$, as a function of $B$.
		Red (big) circles represent the pairs $(B_1, R_{\mathrm{sum}}^\FB(B_1))$ in which $R_{\mathrm{sum}}^\FB(B_1)$ is the information sum-capacity with feedback when only information transmission is performed and $B_1\eqdef 1+2 (1+\rho^\star(1,1)) \SNR$ represents the corresponding maximum energy rate that can be guaranteed at the EH. Blue triangles represent the pairs $(B_\NF,R_{\mathrm{sum}}^\NF(B_\NF))$ in which $R_{\mathrm{sum}}^\NF(B_\NF)$ is the information sum-capacity without feedback and $B_\NF\eqdef 1+2 \SNR$ is the corresponding maximum energy rate that can be guaranteed at the EH without feedback. Orange squares 
		represent the pairs $(B_\F,R_{\mathrm{sum}}^\NF(B_\mathrm{F}))$ in which $B_\F$ is the corresponding maximum energy rate that can be guaranteed at the EH with feedback. Black (small) circles represent the pairs $(B_\mathrm{max},0)$ in which $B_\mathrm{max}\eqdef 1+4 \SNR$ is the maximum energy rate at the EH.}
	\label{FigSumRate}
\end{figure}

In this section, the enhancement on the energy transmission rate due to the  use of feedback is quantified when the information sum-rate is $R_{\mathrm{sum}}^\NF(0)$ (see the blue triangles and orange squares in Fig.~\ref{FigSumRate}).

Denote by $B_\NF = 1 + \SNR_{21} + \SNR_{22}$ the maximum energy rate that can be guaranteed at the EH in the \GMAC$(0)$ when the information sum-rate is $R_{\mathrm{sum}}^\NF(0)$.  Denote also by  $B_{\F}$ the maximum energy rate that can be guaranteed at the EH in the \GMACF$(0)$ when the information sum-rate is $R_{\mathrm{sum}}^\NF(0)$. 
The exact value of $B_{\FB}$ is the solution to an  optimization problem of the form
\begin{IEEEeqnarray}{cCl}
	\nonumber
	B_{\FB} 
	&=&\max\quad B\\
	\text{subject to:}&& R_{\mathrm{sum}}^\FB(B)= R_{\mathrm{sum}}^\NF(0).\label{EqBFB}	
\end{IEEEeqnarray}

The solution to \eqref{EqBFB} is given by the following theorem.
\begin{theorem}\label{LemmaEnergyRateAtSCWF}
	The maximum energy rate $B_{\F}$ that can be guaranteed at the EH in the \GMACF$(0)$ when the information sum-rate is $R_{\mathrm{sum}}^\NF(0)$  is
	\begin{equation}
	\label{EqBLemma}
	B_{\F} = 1 + \SNR_{21} + \SNR_{22} + 2\sqrt{ (1-\gamma) \SNR_{21} \SNR_{22}},
	\end{equation}
	with $\gamma \in (0,1)$ defined as follows:
	\begin{IEEEeqnarray}{lcl}
		\gamma
		& \eqdef & \frac{\SNR_{11} + \SNR_{12}}{2\SNR_{11}\SNR_{12}}\left[ \sqrt{ 1+ \frac{4 \SNR_{11} \SNR_{12}}{\SNR_{11}+\SNR_{12}}}-1\right].\label{eqGamma}
	\end{IEEEeqnarray}
\end{theorem}
\begin{IEEEproof}
	The proof of Theorem~\ref{LemmaEnergyRateAtSCWF} is presented in Appendix~\ref{ProofLemmaEnergyRateAtSCWF}.\end{IEEEproof}

To quantify the energy rate enhancement induced by feedback,  it is of interest to consider the ratio $\frac{B_\F}{B_\NF}$ given by
\begin{IEEEeqnarray}{rCl}
	\label{eq:43}
	\frac{B_\F}{B_{\NF}}&=& 1+\frac{ 2\sqrt{(1-\gamma) \SNR_{21} \SNR_{22}  } }{1+\SNR_{21}+\SNR_{22}}.
\end{IEEEeqnarray}	
Note that the impact of the SNRs in the information transmission branch ($\SNR_{11}$ and $\SNR_{12}$) are captured by $\gamma$.

Let $\nu_i\eqdef \frac{\SNR_{1i}}{\SNR_{1j}}\in \Reals_+$ and $\eta_i\eqdef\frac{\SNR_{2i}}{\SNR_{2j}}\in \Reals_+$, with $(i,j)\in\{1,2\}^2$ and $i\neq j$ measure the asymmetry in the channel from the transmitters to the receiver and to the EH, respectively. Let also $\psi_i\eqdef \frac{\SNR_{2i}}{\SNR_{1i}}\in \Reals_+$ capture the strength ratio between the information and the energy channels of transmitter~$i$.

With these parameters, $\gamma$ in \eqref{eqGamma} can be rewritten as
\begin{equation}
\label{eq:GammaNui}
\gamma=\frac{1+\nu_i}{2 \nu_i \SNR_{1j}} \left[ \sqrt{1+\frac{4 \nu_i \SNR_{1j} }{1+ \nu_i}}-1 \right], \quad \end{equation}
with $(i,j)\in\{1,2\}^2$ and $i\neq j$.

Note that, for all $(i,j)\in \{1,2\}^2$ with $i\neq j$,  when $\SNR_{1j}\to 0$ while the ratio $\nu_i$ remains constant,  from \eqref{eq:GammaNui}, it follows that
\begin{IEEEeqnarray}{rCl}
	\lim_{\SNR_{1j}\to 0}\gamma&=& 1.
\end{IEEEeqnarray}
Thus, when the SNRs in the information branch ($\SNR_{11}$ and $\SNR_{12}$) are very low, the improvement on the energy transmission rate due to feedback is inexistent. This observation is independent of the SNRs in the EH branch ($\SNR_{21}$ and $\SNR_{22}$). 

Alternatively,  when $\SNR_{1j}\to \infty$  while the ratio $\nu_i$ remains constant,  
it follows that 
\begin{IEEEeqnarray}{rCl}
	\lim_{\SNR_{1j}\to \infty}\gamma&=&  0.
\end{IEEEeqnarray}  
Thus, when the SNRs in the information branch ($\SNR_{11}$ and $\SNR_{12}$) are very high, the improvement on the energy transmission rate due to feedback is given by
\begin{equation}
\lim_{\SNR_{1j}\to \infty} \frac{B_\FB}{B_{\NF}}= 1+\frac{2 \sqrt{ \SNR_{21} \SNR_{22}  } }{1+\SNR_{21}+\SNR_{22}}.
\end{equation}

More generally, using the above parameters, the ratio $\frac{B_\F}{B_\NF}$ in \eqref{eq:43} can be written as
\begin{equation}
\frac{B_\F}{B_{\NF}}\hspace*{-1mm}
=\hspace*{-1mm}1\hspace*{-1mm}+\hspace*{-1mm}\frac{2 \psi_j \SNR_{1j} \sqrt{\hspace*{-.5mm}\eta_i \hspace*{-1mm} \left(\hspace*{-1mm}1\hspace*{-1mm}-\hspace*{-1mm}\left(\hspace*{-1mm}\frac{1+\nu_i}{2 \nu_i \SNR_{1j}}\hspace*{-1mm}\left(\hspace*{-1.5mm}\sqrt{\hspace*{-.5mm}1\hspace*{-1mm}+\hspace*{-1mm}\frac{4 \nu_i \SNR_{1j} }{1+ \nu_i}}\hspace*{-1mm}-\hspace*{-1mm}1\hspace*{-1mm} \right)\hspace*{-1mm}\right)\hspace*{-1mm}\right)\hspace*{-1mm}} }{1+(1+\eta_i) \psi_j \SNR_{1j}}.\label{eq:80}
\end{equation}

Based on \eqref{eq:80}, the following corollary evaluates the very low SNR asymptotic energy enhancement with feedback. 	
\begin{corollary}
	\label{Cor:EnergyLowSNR}
	For all $(i,j)\in \{1,2\}^2$ with $i\neq j$,  when $\SNR_{1j}\to 0$ while the ratios $\nu_i, \eta_i,$ and $\psi_i$ remain constant, it holds that
	\begin{equation}
	\lim_{\SNR_{1j}\to 0} \frac{B_\F}{B_{\NF}}=1,
	\end{equation}
	and thus feedback does not enhance energy transmission at very low SNR.
\end{corollary}

In the very high SNR regime, the asymptotic energy enhancement with feedback is given by the following corollary that is also based on \eqref{eq:80}.
\begin{corollary}
	\label{Cor:EnergyHighSNR}
	For all $(i,j)\in \{1,2\}^2$ with $i\neq j$,  when $\SNR_{1j}\to \infty$ while the ratios $\nu_i, \eta_i,$ and $\psi_i$ remain constant, the maximum energy rate improvement with feedback is given by 
	\begin{equation}
	\lim_{\SNR_{1j}\to \infty} \frac{B_\F}{B_{\NF}}=1+\frac{2 \sqrt{\eta_i}}{1+\eta_i}.
	\label{eq:82}
	\end{equation}
\end{corollary}
From Corollary~\ref{Cor:EnergyLowSNR} and Corollary~\ref{Cor:EnergyHighSNR}, it holds that:
\begin{corollary}
	Feedback can at most double the energy transmission rate:
	\begin{equation}
	1 \leqslant \frac{B_{\F}}{B_{\mathrm{NF}}} \leqslant 2,
	\end{equation}
	where the upper-bound holds with equality when $\eta_i=1$, i.e., $\SNR_{21}=\SNR_{22}$.
\end{corollary}

\begin{figure}[t]
	\centering
	\includegraphics[scale=0.43]{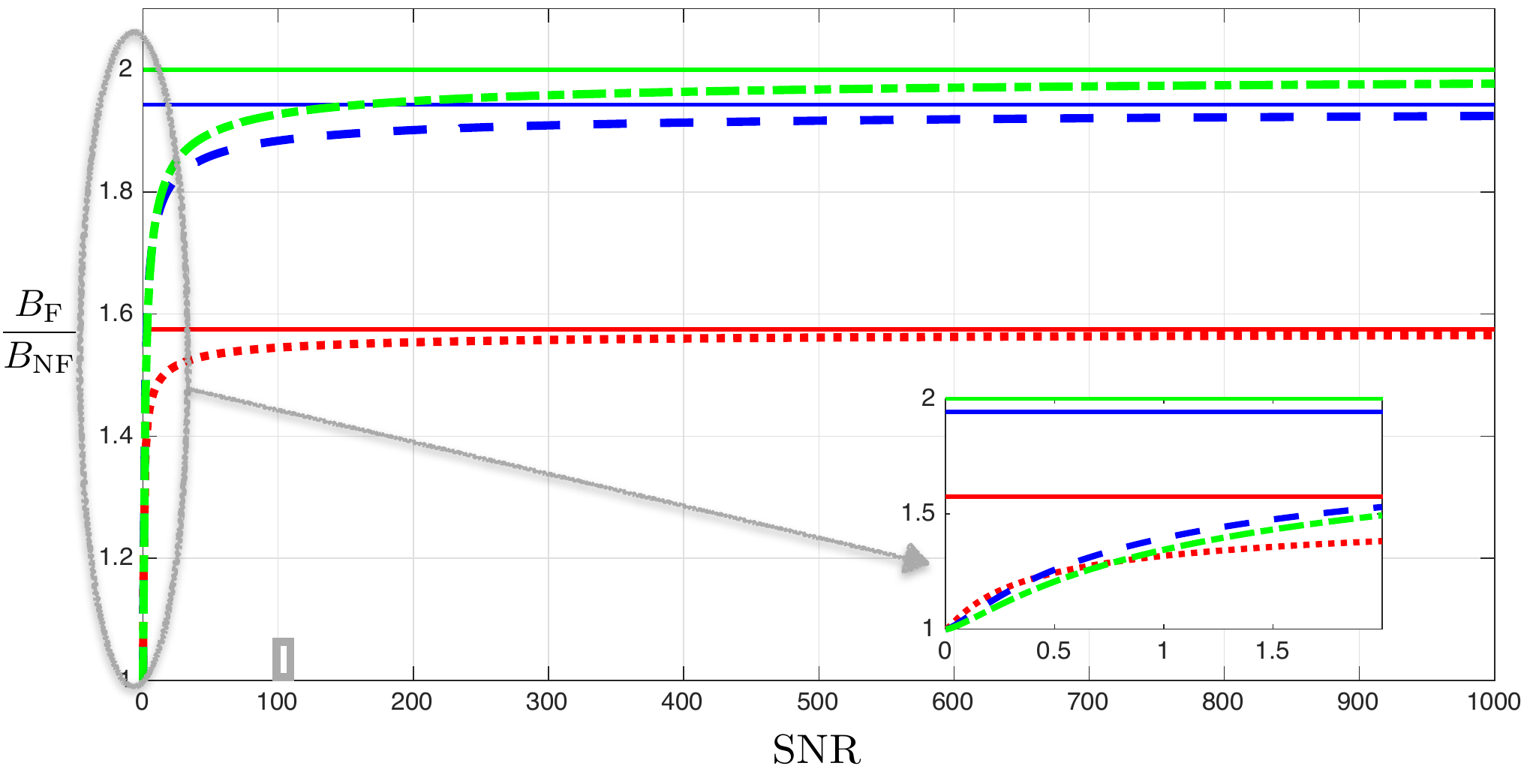}
	\caption{The ratio  $\frac{B_{\F}}{B_{\mathrm{NF}}}$  and its high-SNR limit as a function of $\SNR$ when the receiver and the EH are co-located and  $\SNR_{11}=\SNR_{21}=\SNR_1$ and $\SNR_{12}=\SNR_{22}=\SNR_2$. 
		The solid line is the high-SNR limit  in \eqref{eq:82}; the dash-dotted line, the dashed line and the dotted line are the exact values of the ratio $\frac{B_{\F}}{B_{\mathrm{NF}}}$ in \eqref{eq:80} when $\SNR_1 = \SNR_2 = \SNR$;  $\frac{\SNR_1}{2} =  \SNR_2 = \SNR$; and $\frac{\SNR_1}{10} =  \SNR_2 = \SNR$, respectively. }
	\label{FigRatio}
\end{figure}

Fig.~\ref{FigRatio} compares the exact value of the ratio $\frac{B_\F}{B_\NF}$ in \eqref{eq:80} to the high-SNR limit in \eqref{eq:82} as a function of the SNRs in the special case in which the receiver and the EH are co-located. 	This implies that the channel coefficients between the transmitters and the receiver are identical to those between the transmitters and the EH., i.e., $\SNR_{11}=\SNR_{21}=\SNR_1$ and $\SNR_{12}=\SNR_{22}=\SNR_2$. Note that in the symmetric case, i.e., $\SNR_1=\SNR_2=\SNR$, the upper-bound in  \eqref{eq:82} is tight since the ratio $ \frac{B_\F}{B_\NF}$ becomes arbitrarily close to two as $\SNR$ tends to infinity.  
In the non-symmetric cases $\SNR_1 \neq \SNR_2$, this bound is loose. 

\section{Conclusion and Extensions}
\label{SecConclusion}
This report has characterized the information-energy capacity region of the two-user G-MAC with an EH, with and without feedback, and has determined the energy transmission enhancement induced by the use of feedback. An important conclusion of this work
is that SEIT requires additional transmitter cooperation/coordination.  
From this viewpoint, any technique  that allows transmitter cooperation (i.e., feedback, conferencing, etc.) is likely to provide performance gains in SEIT in general multi-user networks. The results on the energy transmission enhancement induced by feedback in the two-user G-MAC-F can be extended to the $K$-user G-MAC-F with EH for arbitrary $K\geqslant 3$.   

\appendix
\section{Proof of Theorem \ref{Theorem-EC-Region-MAC-F}}
\label{SecProofThm3}
The proof is divided into two parts: achievability and converse parts.
\subsection{Proof of Achievability}\label{SecAchievability}

The proof of achievability uses a very simple power-splitting technique in which a fraction $\beta_i\in [0,1]$ of the power is used for information transmission and the remaining fraction $(1-\beta_i)$  for energy transmission. The information transmission is made following Ozarow's perfect feedback capacity-achieving scheme in~\cite{Ozarow-TIT-1984}. The energy transmission is accomplished by random symbols that are known at both transmitters and the receiver. Despite a great deal of similarity with the scheme in~\cite{Ozarow-TIT-1984}, the complete proof is fully described hereunder for the sake of completeness. 

\textbf{Codebook generation:}  At the beginning of the transmission, each message $M_i$ is mapped  into the real-valued message point 
\begin{equation}\label{eq:Theta}
\Theta_{i}(M_i)\triangleq- (M_i-1)\Delta_i + \sqrt{P_i},
\end{equation}  
where \begin{equation}
\label{eq:Delta}
\Delta_i\triangleq \frac{2\sqrt{P_i}}{\lfloor2^{ nR_i}\rfloor}.\end{equation}

\textbf{Encoding:} The first three channel uses are part of an initialization procedure during which there is no energy transmission and the channel inputs are
\begin{subequations}\label{eq:MACinit_inputs}
	\begin{IEEEeqnarray}{llll}
		t=-2:&\hspace{0.2cm}& X_{1,-2}=0&\quad \text{and}\quad X_{2,-2}=\Theta_2(M_2),\IEEEeqnarraynumspace\\  
		t=-1:&& X_{1,-1}=\Theta_1(M_1)&\quad \text{and}\quad X_{2,-1}=0,\\
		t=0:&& X_{1,0}=0&\quad \text{and}\quad X_{2,0}=0.
	\end{IEEEeqnarray}
\end{subequations}

Through the feedback links, transmitter~1 observes $(Z_{-1},Z_{0})$ and transmitter~2 observes $(Z_{-2},Z_{0})$.  After the initialization phase, each transmitter~$i\in\{1,2\}$ can thus compute  
\begin{IEEEeqnarray}{rCl}\label{eq:Xi}
	\Xi_i&\eqdef&\sqrt{1-\rho^\star(\beta_1,\beta_2)}\cdot Z_{-i}+ \sqrt{\rho^\star(\beta_1,\beta_2)}\cdot  Z_0,
\end{IEEEeqnarray}
where $\rho^\star(\beta_1,\beta_2)$ is the unique solution in $(0,1)$ to \eqref{eq:rhostar}. 

During the remaining channel uses $1,\ldots, n$, for $i \in \{1,2\}$, instead of repeating the message-point $\Theta_i(M_i)$, transmitter~$i$  simultaneously describes  $\Xi_i$ to the receiver and transmits energy to the EH. Let $\beta_i$, with $i \in \lbrace 1, 2\rbrace$ be the power-splitting coefficient of transmitter $i$. More specifically, at each time $t\in \{1,\dots,n\}$, transmitter~$i$ sends
\begin{IEEEeqnarray}{rCl}
	X_{i,t}&=& U_{i,t}+\sqrt{(1-\beta_i) P_i} W_t, \quad i \in \{1,2\}.
\end{IEEEeqnarray}
Here $(W_1,\dots,W_n)$ is an independent and identically distributed (i.i.d.) sequence drawn according to a zero-mean unit-variance Gaussian distribution. This sequence is known non-causally to the transmitters and to the receiver and is independent of the messages and  the noise sequences. The symbol $U_{i,t}$ is a zero-mean Gaussian random variable with  variance $\beta_i P_i$ and is chosen as follows:
\begin{subequations}
	\label{eq:uit}
	\begin{IEEEeqnarray}{rCl}
		U_{i,1}&=&\sqrt{\beta_i P_i}\;\Xi_{i},\\
		U_{i,t}&=& \gamma_{i,t} \biggr( \Xi_{i}- \hat{\Xi}_{i}^{(t-1)} \biggr), \quad t \in \{2,\dots,n\},
	\end{IEEEeqnarray}
\end{subequations}
where the parameter $\gamma_{i,t}$ is chosen to satisfy 
$\E{U_{i,t}^2}=\beta_i P_i$ and $\hat{\Xi}_{i}^{(t-1)}$ is explained below. 

For each $t\in \{1,\dots,n\}$, upon receiving the channel output $Y_{1,t}$, 
the receiver subtracts the signal induced by the common randomness
to form the observation $Y'_{1,t}$ as follows:
\begin{equation}
Y'_{1,t}\eqdef Y_{1,t}-\left(h_{11}\sqrt{(1-\beta_1)P_1}+h_{12}\sqrt{(1-\beta_2)P_2}\right)W_t.
\end{equation}
The receiver then calculates the minimum mean square error (MMSE) estimate\break $\hat{\Xi}_{i}^{(t-1)}=\E{\Xi_{i}|Y'_{1,1},\dots,Y'_{1,t-1}}$  of $\Xi_i$ given the prior observations $Y'_{1,1},\dots,Y'_{1,t-1}$. 

\begin{remark}
	Note that by the orthogonality principle of MMSE estimation~\cite{KailathSayedHassibi}, $(U_{1,t},U_{2,t},Z_t)$ are independent of the  observations $Y'_{1,1},\dots,Y'_{1,t-1}$ and thus of $Y_{1,1},\dots,Y_{1,t-1}$. Furthermore, since $(W_1,\dots,W_n)$ are i.i.d., it holds that, for any $i\in \{1,2\}$ and for any $t\in \{1,\dots,n\}$, $Y_{i,t}$ is independent of $Y_{i,1},\dots,Y_{i,t-1}$.
\end{remark}
\begin{remark}
	Let $\rho_t$ denote the correlation coefficient between $U_{1,t}$ and $U_{2,t}$, i.e., $\rho_t\hspace*{-1mm}\eqdef\hspace*{-1mm}\frac{\E{U_{1,t} U_{2,t} }}{\sqrt{\E{U_{1,t}^2 }\E{U_{2,t}^2 }}}$. In \cite[Lemma 17.1]{NIT}, it is proved that for all $t\in \{1,\dots,n\}$, $\rho_t=\rho^\star(\beta_1,\beta_2)$, and thus $\rho^\star(\beta_1,\beta_2)$ is the steady-state correlation coefficient.
\end{remark}
After reception of the output symbols $Y_{1,-2},\ldots,Y_{1,n}$, the receiver forms\break $\hat{\Xi}_{i}^{(n)}\eqdef \E{\Xi_i|Y'_{1,1},\dots,Y'_{1,n}},$ for $i\in\{1,2\}$. Then, it forms an estimate $\hat{\Theta}_i^{(n)}$ of the  message point $\Theta_i(M_i)$ as follows:
\begin{IEEEeqnarray}{rCl}
	\hat{\Theta}_i^{(n)}\hspace*{-1mm} &\eqdef &\hspace*{-1mm} \frac{1}{h_{1i}}\hspace*{-1mm} \Biggr(\hspace*{-1mm}
	Y_{1,-i}\hspace*{-0.5mm}+\hspace*{-0.5mm}\sqrt{\hspace*{-1mm}\frac{ \rho^\star(\beta_1,\beta_2)}{1-\rho^\star(\beta_1,\beta_2)}} Y_{1,0}\hspace*{-0.5mm}-\hspace*{-0.5mm}  \frac{1}{\sqrt{1-\rho^\star(\beta_1,\beta_2)}}\hat{\Xi}_{i}^{(n)}\hspace*{-1mm}\Biggr)\hspace*{-1mm} \nonumber\\
	&=& 
	\Theta_i(M_i)+ \frac{1}{h_{1i}\sqrt{1-\rho^\star(\beta_1,\beta_2)}} \left(\Xi_{i}-\hat{\Xi}_{i}^{(n)}\right).
\end{IEEEeqnarray}

Finally, the message index estimate $M_i$ is obtained using nearest-neighbor decoding based on the value $\hat{\Theta}_i^{(n)}$, as follows: 
\begin{equation}\label{eq:nearest}
\hat{M}_i^{(n)} = \argmin_{m_i\in\{1,\ldots,  \lfloor 2^{nR_i}\rfloor\}} \big|\Theta_i(m_i)- \hat{\Theta}_i^{(n)}\big|.
\end{equation}

\textbf{Analysis of the probability of error:}

An error occurs whenever the receiver is not able to recover one of the messages, i.e., $(M_1,M_2)\neq (\hat{M}_1^{(n)},\hat{M}_2^{(n)})$ or if the received energy rate is below the desired minimum rate $B^{(n)}<B$.

First, consider the probability of a decoding error. Note that for $i\in\{1,2\}$,  $\hat{M}_i^{(n)}=M_i$, if 
\begin{equation}
|\Xi_{i}- \hat{\Xi}_{i}^{(n)}|\leqslant \frac{h_{1i} \sqrt{1-\rho^\star(\beta_1,\beta_2)} \Delta_i}{2}.
\end{equation}

Since the difference $\Xi_{i}-\hat{\Xi}_{i}^{(n)}$ is a centered Gaussian random variable, by the definition of $\Delta_i$ in \eqref{eq:Delta}, the error probability $P_{e,i}^{(n)}$ while decoding message index $M_i$ can be bounded as 
\begin{IEEEeqnarray}{rCl}
	P_{e,i}^{(n)} \leqslant 2\Q \left(   \frac{\sqrt{\SNR_{1i}}  \sqrt{1-\rho^\star(\beta_1,\beta_2)} }{\lfloor 2^{nR_i} \rfloor \sqrt{(\sigma_{i}^{(n)})^2}} \right),\label{eq:Pe}
\end{IEEEeqnarray}
where $\Q(x)=\frac{1}{\sqrt{2\pi}} \int_x^\infty \exp\left(-\frac{u^2}{2}\right) du$ is the tail of the unit Gaussian distribution evaluated at $x$ and where
\begin{IEEEeqnarray}{rCl} 
	(\sigma_{i}^{(n)})^2&\eqdef&\E{|\Xi_{i}-\hat{\Xi}_{i}^{(n)}|^2}, \quad i\in \{1,2\}.
\end{IEEEeqnarray}

Note that
\begin{IEEEeqnarray}{rCl}
	I(\Xi_{i}; \vect{Y}'_1) & = & h(\Xi_{i}) - h(\Xi_{i}|\vect{Y}'_1)\nonumber \\
	& \stackrel{(a)}{=} &  h(\Xi_{i}) - h(\Xi_{i}-\hat{\Xi}_{i}^{(n)} |\vect{Y}'_1)\nonumber\\  
	& \stackrel{(b)}{=} &  h(\Xi_{i})- h(\Xi_{i}-\hat{\Xi}_{i}^{(n)})\nonumber\\
	& = & -\frac{1}{2} \log_2 \bigr(  (\sigma_{i}^{(n)})^2\bigr),\label{eq:ii}
\end{IEEEeqnarray}
where $(a)$ holds because by the joint Gaussianity of $\Xi_{i}$ and $\vect{Y}'_1$, the MMSE estimate $\hat{\Xi}_{i}^{(n)}$ is a linear function of $\vect{Y}'_1$ (see, e.g.,~\cite{LapidothBook}); $(b)$ follows because by the orthogonality principle, the error $\Xi_{i}-\hat{\Xi}_{i}^{(n)}$ is independent of the observations $\vect{Y}'_1$. 

Equation~\eqref{eq:ii} can equivalently be rewritten as
\begin{equation}
\sqrt{(\sigma_{i}^{(n)})^2}=2^{- I(\Xi_{i}; \vect{Y}'_1)}.\label{eq:iii} 
\end{equation}

Combining~\eqref{eq:Pe} with \eqref{eq:iii} yields that the probability of error of message $M_i$ tends to 0 as $n\to \infty$, if the rate $R_i$ satisfies
\begin{IEEEeqnarray}{rCl}
	R_{i} &\leqslant &\liminf_{n\to\infty} \frac{1}{n} I(\Xi_{i};  \vect{Y}'_1),\qquad i\in\{1,2\}.\label{eq:liminf}
\end{IEEEeqnarray}

On the other hand, as proved in \cite[Sec.~17.2.4]{NIT}, 
\begin{IEEEeqnarray}{rCl}
	I(\Xi_{i};\vect{Y}_1') & = &\sum_{t=1}^n I(U_{i,t};Y'_{1,t})
\end{IEEEeqnarray}
and irrespective of $n$ and $t\in\{1,\ldots, n\}$, it holds that 
\begin{IEEEeqnarray}{rCl}
	I(U_{i,t};Y'_{1,t}) & =&\frac12 \log_2\left(1+\beta_i\SNR_{1i} (1-(\rho^\star(\beta_1,\beta_2))^2)\right).\IEEEeqnarraynumspace
\end{IEEEeqnarray}

Hence, for $i\in\{1,2\}$ it holds that
\begin{IEEEeqnarray}{rCl}
		\liminf_{n\to\infty} \frac{1}{n} I(\Xi_{i};  \vect{Y}'_1)
	&=&\frac12 \log_2\left(1+\beta_i\SNR_{1i} (1-(\rho^\star(\beta_1,\beta_2))^2)\right).\label{eq:RiOzlim}
\end{IEEEeqnarray}

Combining \eqref{eq:liminf} and \eqref{eq:RiOzlim} yields that when $n \to \infty$, this scheme can achieve all non-negative rate-pairs $(R_1,R_2)$ that satisfy
\begin{subequations}
	\begin{IEEEeqnarray}{cCl}
		R_1&\leqslant & \frac12 \log_2\left(1+\beta_1\SNR_{11} (1-{\rho^\star(\beta_1,\beta_2)}^2)\right),\\
		R_2&\leqslant & \frac12 \log_2\left(1+\beta_2\SNR_{12} (1-{\rho^\star(\beta_1,\beta_2)}^2)\right).\IEEEeqnarraynumspace
	\end{IEEEeqnarray}
	Hence, combined with $\eqref{eq:rhostar}$, it automatically yields
	\begin{multline}
	R_1+R_2 \leqslant \frac12 \log_2\Bigr(1+\beta_1\; \SNR_{11}+\beta_2\;\SNR_{12}+2\rho^\star(\beta_1,\beta_2)\sqrt{\beta_1\SNR_{11} \beta_2\SNR_{12}} \Bigr).
	\end{multline}
\end{subequations}

Furthermore, the total consumed power at transmitter~$i$ for $i\in \{1,2\}$ over the $n+3$ channel uses is upper bounded by $(n+1) P_i$, hence, this scheme satisfies the input-power constraints.

\textbf{Average received energy rate:}

The average received energy rate is given by ${B^{(n)}\eqdef\frac1n \sum_{t=1}^n Y_{2,t}^2}$.

By the memoryless property of the channel and by the choice of the inputs, the sequence $Y_{2,1},\dots,Y_{2,n}$ is i.i.d. and each $Y_{2,t}$ follows a zero-mean Gaussian distribution with variance $\bar B$ given by
\begin{IEEEeqnarray}{lcl}
	\bar B&\eqdef &\E{Y_{2,t}^2}\nonumber\\
	&=& 1 + \SNR_{21} + \SNR_{22} + 2 \sqrt{\beta_1 \SNR_{21} \beta_2\SNR_{22}}\rho^\star(\beta_1,\beta_2)+2\sqrt{(1-\beta_1)\SNR_{21}(1-\beta_2)\SNR_{22}},\nonumber\\
\end{IEEEeqnarray} 
where the correlation among the IC components is in the steady state.

By the weak law of large numbers,  it holds that
$\forall \epsilon >0$,
\begin{IEEEeqnarray}{rCl}
	\lim_{n\to\infty} \Pr\left(|B^{(n)}- \bar B|>\epsilon\right) =0.
\end{IEEEeqnarray}
Consequently,
\begin{subequations}
	\begin{IEEEeqnarray}{l}
		\lim_{n\to\infty} \Pr\left(B^{(n)} >\bar B+\epsilon\right) =0,\quad \text{and}\\	
		\lim_{n\to\infty} \Pr\left(B^{(n)} <\bar B-\epsilon\right) =0.\label{eq:limBmax}
	\end{IEEEeqnarray}
\end{subequations}
From~\eqref{eq:limBmax}, it holds that for any energy rate $B$ which satisfies  $0<B\leqslant \bar B$, it holds that
\begin{IEEEeqnarray}{rCl}
	\lim_{n\to\infty} \Pr\left(B^{(n)} <B-\epsilon\right) =0.
\end{IEEEeqnarray}

To sum up, any information-energy rate triplet $(R_1,R_2,B)$ that satisfies
\begin{subequations}
	\begin{IEEEeqnarray}{cCl}
		R_1&\leqslant& \frac12 \log_2\left(1+\beta_1\SNR_{11} (1-{(\rho^\star(\beta_1,\beta_2))}^2)\right)\label{eq:cr1}\IEEEeqnarraynumspace\\
		R_2&\leqslant & \frac12 \log_2\left(1+\beta_2\SNR_{12} (1-{(\rho^\star(\beta_1,\beta_2))}^2)\right)\\
		R_1+R_2&\leqslant &\frac12 \log_2\Bigr(1+\beta_1 \SNR_{11}+\beta_2\SNR_{12}+2\rho^\star(\beta_1,\beta_2)\sqrt{\beta_1\;\SNR_{11} \cdot \beta_2\;\SNR_{12}} \Bigr)\label{eq:cr12}\\
		B&\leqslant &1 + \SNR_{21} + \SNR_{22} + 2 \Bigr(\sqrt{\beta_1 \beta_2}\rho^\star(\beta_1,\beta_2)+\sqrt{(1-\beta_1)(1-\beta_2)}\Bigr) \sqrt{\SNR_{21} \SNR_{22}}\IEEEeqnarraynumspace
	\end{IEEEeqnarray}
	\label{eq:region_star}\end{subequations}
is achievable.

To achieve other points in the  information-energy capacity region, transmitter~1 can split its message $M_1$ into two independent submessages $(M_{1,0},M_{1,1})\in \{1,\dots,\lfloor 2^{nR_{1,0}}\rfloor\}\times \{1,\dots,\lfloor2^{nR_{1,1}}\rfloor\}$ such that $R_{1,0},R_{1,1}\geq0$ and $R_{1,0}+R_{1,1}=R_1$.
It uses a power fraction $\alpha_1\in [0,1]$ of its available information-dedicated power $\beta_1 P_1$ to transmit $M_{1,0}$ using a non-feedback Gaussian random code  and uses the remaining power $(1-\alpha_1) \beta_1 P_1$ to send $M_{1,1}$ using the sum-capacity-achieving feedback scheme while treating $M_{1,0}$ as noise. Transmitter~$2$ sends its message $M_2$ using the sum-capacity-achieving feedback scheme. 

Transmitter~$1$'s IC-input is $U_{1,t}\eqdef U_{1,0,t}+U_{1,1,t}$ where $U_{1,1,t}$ is defined as in \eqref{eq:uit} but with reduced power ${(1-\alpha_1) \beta_1 P_1}$, and $U_{1,0,t}$ is an independent zero-mean Gaussian random variable with variance $\alpha_1 \beta_1 P_1$.  Transmitter~$2$'s IC-input is defined as in \eqref{eq:uit}.

The receiver first subtracts the common randomness and then decodes $(M_{1,1},M_2)$ treating the signal encoding $M_{1,0}$ as noise. Successful decoding is possible if
\begin{subequations}
	\begin{IEEEeqnarray}{rCl}
		R_{1,1}\hspace*{-1mm}&\leqslant&\hspace*{-1mm}\frac12 \log_2\hspace*{-1mm}\left(\hspace*{-1mm}1+\frac{(1-\alpha_1) \beta_1\SNR_{11} (1-{\rho_{\alpha_1}(\beta_1,\beta_2)}^2)}{1+\alpha_1 \beta_1 \SNR_{11} }\hspace*{-0.5mm}\right)\hspace*{-2mm}\IEEEeqnarraynumspace\\
		R_2\hspace*{-1mm}&\leqslant&\hspace*{-1mm}\frac12 \log_2\left(1+\frac{\beta_2 \SNR_{12} (1-{\rho_{\alpha_1}(\beta_1,\beta_2)}^2)}{1+\alpha_1 \beta_1 \SNR_{11}}\right),\hspace*{-2mm}
	\end{IEEEeqnarray}
\end{subequations}
where $\rho_{\alpha_1}(\beta_1,\beta_2)$ is defined as follows. When $\beta_1\neq 0$, $\beta_2\neq 0$, and $\alpha_1\neq 1$, $\rho_{\alpha_1}(\beta_1,\beta_2)$
is the unique solution in $(0,1)$ to the following equation in $x$:
\begin{IEEEeqnarray}{rCl}\label{eq:rhostaralpha}
	\lefteqn{\hspace*{-3mm}
		1\hspace*{-1mm}+\hspace*{-1mm}\frac{(1\hspace*{-1mm}-\hspace*{-1mm}\alpha_1)\beta_1 \SNR_{11}\hspace*{-1mm}+\hspace*{-1mm}\beta_2 \SNR_{12}\hspace*{-1mm}+\hspace*{-1mm}2 x \hspace*{-.5mm}\sqrt{\hspace*{-1mm}\beta_1 \beta_2\hspace*{-.5mm}(1\hspace*{-1mm}-\hspace*{-1mm}\alpha_1) \SNR_{11} \SNR_{12}} }{1+\alpha_1 \beta_1 \SNR_{11}} }\nonumber\\
	&\hspace*{5mm}=\hspace*{-1mm}&\left(\hspace*{-1mm}1\hspace*{-1mm}+\hspace*{-1mm}\frac{(1\hspace*{-1mm}-\hspace*{-1mm}\alpha_1) \beta_1\SNR_{11}}{1\hspace*{-1mm}+\hspace*{-1mm}\alpha_1 \beta_1 \SNR_{11}}\hspace*{-1mm}(1\hspace*{-1mm}- x^2)\hspace*{-1mm}\right)\hspace*{-1.5mm}\left(\hspace*{-1mm}1\hspace*{-1mm}+\hspace*{-1mm}\frac{\beta_2 \SNR_{12}}{1\hspace*{-1mm}+\hspace*{-1mm}\alpha_1 \beta_1 \SNR_{11}} (1\hspace*{-1mm}-\hspace*{-1mm} x^2\hspace*{-.5mm})\hspace*{-1mm}\right)\hspace*{-1mm},
\end{IEEEeqnarray}
In this case, the existence and the uniqueness of $ \rho_{\alpha_1}(\beta_1,\beta_2)$ follow a similar argument as the existence and uniqueness of a solution to~\eqref{eq:rhostar}.
When $\alpha_1= 1$,
$\rho_{\alpha_1}(\beta_1,\beta_2)=\rho^\star(\beta_1,\beta_2)$. When either $\beta_1= 0$ or $\beta_2= 0$, regardless of the value of $\alpha_1$, $\rho_{\alpha_1}(\beta_1,\beta_2)=0$.

Then, using successive interference cancellation, the receiver recovers $M_{1,0}$ successfully if
\begin{IEEEeqnarray}{rCl}
	R_{1,0}&\leqslant & \frac12 \log_2\left(1+ \alpha_1 \beta_1 \SNR_{11} \right).
\end{IEEEeqnarray}

By substituting $R_1=R_{1,0}+R_{1,1}$, it can be seen that successful decoding of $(M_1,M_2)$ is possible with arbitrarily small probability of error if the rates $(R_1,R_2)$ satisfy 
\begin{subequations}
	\label{eq:const_oz_alpha}
	\begin{IEEEeqnarray}{rCl}
		R_{1}&\leqslant &\frac12 \log_2\left(1+\frac{ (1-\alpha_1)\beta_1 \SNR_{11} \bigr(1-{(\rho_{\alpha_1}(\beta_1,\beta_2))}^2\bigr)}{1+\alpha_1 \beta_1 \SNR_{11}}\right)+\frac12 \log_2\left(1+ \alpha_1 \beta_1 \SNR_{11} \right)\\
		R_2&\leqslant &\frac12 \log_2\left(1+\frac{\beta_2 \SNR_{12} \bigr(1-{(\rho_{\alpha_1}(\beta_1,\beta_2))}^2\bigr)}{1+\alpha_1 \beta_1\SNR_{11}}\right).
	\end{IEEEeqnarray}
\end{subequations}

Now, the average received energy rate of this scheme is analyzed. The sequence $Y_{2,1},\dots,Y_{2,n}$ is i.i.d. and each $Y_{2,t}$ for $t\in \{1,\dots,n\}$ follows a zero-mean Gaussian distribution with variance $B$ given by
\begin{multline}
B=\E{Y_{2,t}^2}	=1 + \SNR_{21} + \SNR_{22}+ 2 \sqrt{1-\alpha_1}\;\rho_{\alpha_1}(\beta_1,\beta_2)\sqrt{ \beta_1\SNR_{21} \beta_2\SNR_{22}}\\+2\sqrt{(1-\beta_1)\SNR_{21}(1-\beta_2)\SNR_{22}}\bigr).\label{eq:Bconst}
\end{multline}

Here also the weak law of large numbers implies that 
\begin{IEEEeqnarray}{rCl}
	\lim_{n\to\infty} \Pr\left(B^{(n)} <b-\epsilon\right) =0
\end{IEEEeqnarray}
for any $b\in [0,B]$. 

Now if  $\rho$ replaces  ${\sqrt{1-\alpha_1 }\;\rho_{\alpha_1}(\beta_1,\beta_2)}$ with $\alpha_1\in [0,1]$ in constraints~\eqref{eq:const_oz_alpha} and~\eqref{eq:Bconst},  then  any non-negative information-energy rate triplet $(R_1,R_2,B)$ satisfying
\begin{subequations}
	\label{eq:region1oz}
	\begin{IEEEeqnarray}{rCl}
		R_1&\leqslant & \frac12 \hspace*{-.5mm}\log_2\hspace*{-1mm}\left(1\hspace*{-1mm}+\hspace*{-1mm}\beta_1\SNR_{11} \left(1-{\rho}^2\right)\right),\\
		R_2&\leqslant& \frac12\hspace*{-.5mm} \log_2\hspace*{-1mm}\left(\hspace*{-1mm}1\hspace*{-1mm}+\hspace*{-1mm}\beta_1 \SNR_{11}\hspace*{-1mm}+\hspace*{-1mm}\beta_2 \SNR_{12}\hspace*{-1mm}+\hspace*{-1mm}2\rho\sqrt{\beta_1 \SNR_{11}\beta_2\SNR_{12}}\hspace*{-.5mm}\right)
		-\frac12 \log_2\left(1\hspace*{-1mm}+\hspace*{-1mm}\beta_1 \SNR_{11} \left(1-\rho^2\right)\right)\hspace*{-1mm},
		\IEEEeqnarraynumspace\\
		B&\leqslant&1\hspace*{-1mm} +\hspace*{-1mm} \SNR_{21}\hspace*{-1mm} +\hspace*{-1mm} \SNR_{22}+ 2\hspace*{-.5mm}\left(\hspace*{-1mm}\rho \sqrt{ \beta_1 \beta_2}\hspace*{-1mm}+\hspace*{-1mm}\sqrt{(1-\beta_1)(1-\beta_2)}\right)\hspace*{-1mm}\sqrt{\SNR_{21} \SNR_{22}},
	\end{IEEEeqnarray}
\end{subequations}
where $\rho\in [0,\rho^\star(\beta_1,\beta_2)]$ and $\rho^\star(\beta_1,\beta_2)$ is the unique solution to~\eqref{eq:rhostar}, is achievable.

If the roles of transmitters $1$ and $2$ are reversed, it can be shown that any non-negative information-energy rate triplet $(R_1,R_2,B)$ such that
\begin{subequations}
	\label{eq:region2oz}
	\begin{IEEEeqnarray}{rCl}
		R_1\hspace*{-1mm}&\leqslant&\hspace*{-1mm}\frac12\hspace*{-0.5mm} \log_2\hspace*{-1mm}\left(\hspace*{-0.8mm}1\hspace*{-0.8mm}+\hspace*{-0.8mm}\beta_1\SNR_{11}\hspace*{-0.8mm}+\hspace*{-0.8mm}\beta_2\SNR_{12}\hspace*{-0.8mm}+\hspace*{-0.8mm}2\rho \sqrt{\hspace*{-0.5mm}\beta_1\SNR_{11}\beta_2 \SNR_{12}}\hspace*{-0.5mm}\right)-\frac12 \log_2\hspace*{-1mm}\left(1+\beta_2\SNR_{12} (1-\rho^2)\right),\IEEEeqnarraynumspace\\
		R_2\hspace*{-1mm}&\leqslant &\hspace*{-1mm} \frac12\hspace*{-0.5mm} \log_2\hspace*{-1mm}\left(1+\beta_2 \SNR_{12} (1-{\rho}^2)\right),\\
		B&\leqslant&\hspace*{-0.5mm}1 + \SNR_{21} + \SNR_{22} + 2 \rho \sqrt{ \beta_1\SNR_{21} \beta_2 \SNR_{22}}+ 2\sqrt{(1-\beta_1)\SNR_{21}(1-\beta_2)\SNR_{22}},
	\end{IEEEeqnarray}
\end{subequations}
for any $\rho\in [0,\rho^\star(\beta_1,\beta_2)]$, is achievable. 

Time-sharing between all information-energy rate triplets in the union of the two regions described by the constraints \eqref{eq:region1oz} and \eqref{eq:region2oz} concludes the proof of achievability of the region. This yields 
\begin{subequations}
	\begin{IEEEeqnarray}{ccl}
		R_1 & \leqslant & \frac{1}{2} \log_2\left( 1 + \beta_1 \SNR_{11}\left( 1 - \rho^2 \right)  \right),\label{eq:R1_oz}\\
		R_2 & \leqslant & \frac{1}{2} \log_2\left( 1 + \beta_2 \SNR_{12} \left( 1 - \rho^2 \right)   \right),\label{eq:R2_oz}\\
		R_1 + R_2 & \leqslant & \frac{1}{2} \log_2\Bigr( 1 + \beta_1 \SNR_{11} + \beta_2 \SNR_{12}+ 2 \rho\sqrt{\beta_1\SNR_{11}\ \beta_2 \SNR_{12}} \Bigr), \label{eq:R12_oz}\IEEEeqnarraynumspace\\
		B  & \leqslant &  1 + \SNR_{21} + \SNR_{22} + 2 \rho \sqrt{ \beta_1\SNR_{21} \beta_2 \SNR_{22}}\nonumber\\&&+ 2\sqrt{(1-\beta_1)\SNR_{21} (1-\beta_2)\SNR_{22}},\label{eq:B_oz}
	\end{IEEEeqnarray}
\end{subequations}
for any $\rho\in [0,\rho^\star(\beta_1,\beta_2)]$.

Note that for any $\rho>\rho^\star(\beta_1,\beta_2)$,  
the sum of \eqref{eq:R1_oz} and \eqref{eq:R2_oz} is strictly smaller than \eqref{eq:R12_oz}. The resulting information  region is a rectangle that is strictly contained in the rectangle obtained for $\rho=\rho^\star(\beta_1,\beta_2)$. In other words, there is no gain in terms of information rates.
In terms of energy rates, for any  $\rho>\rho^\star(\beta_1,\beta_2)$, there always exists a pair $(\beta'_1,\beta'_2)$ such that 
$$\rho=\sqrt{\beta'_1 \beta'_2} \rho^\star(\beta'_1, \beta'_2) + \sqrt{(1-\beta'_1)(1-\beta'_2)}.$$
This choice achieves any information rate pair $(R_1,R_2)$ satisfying
\begin{IEEEeqnarray}{rCl}
	R_i \leqslant \frac12 \log_2\left(1+\beta'_i \SNR_{1i}(1-\rho^\star(\beta'_1,\beta'_2)^2) \right).
\end{IEEEeqnarray}	
In particular, it achieves 
\begin{IEEEeqnarray}{rCl}
	R_i \leqslant \frac12 \log_2\left(1+\beta'_i \SNR_{1i}(1-\rho^2)\right),\quad i \in \{1,2\},
\end{IEEEeqnarray}
since $\rho>\rho^\star(1,1)=\ds\max_{(\beta_1,\beta_2)\in [0,1]^2} \rho^\star(\beta_1,\beta_2)$.	
This completes the proof of the achievability part of Theorem \ref{Theorem-EC-Region-MAC-F}.
\subsection{Proof of Converse}
Fix  an information-energy rate triplet $(R_1,R_2,B)\in\set{E}_b^\FB(\SNR_{11},\SNR_{12},\SNR_{21},\SNR_{22})$. For this information-energy rate triplet and for
each blocklength $n$, encoding and decoding functions are chosen
such that  
\begin{subequations}
	\begin{IEEEeqnarray}{cll}
		\limsup_{n \rightarrow \infty}\;  P_{\error}^{(n)}  & = & 0,\label{assumption1}\\
		\limsup_{n \rightarrow \infty}\;  P_\outage^{(n)}   &=&  0\text{ for any $\epsilon>0$,}\label{assumption2}\\
		B&\geqslant & b, \label{assumption3}
	\end{IEEEeqnarray}
\end{subequations} subject to  the input power constraint~\eqref{EqPowerConstraint}.

Using assumption~\eqref{assumption1}, applying Fano's inequality and following similar steps as in \cite{Ozarow-TIT-1984}, it can be shown that the rate-pair
$(R_1,R_2)$ must satisfy
\begin{subequations}
	\label{eq:sum_outer_bound}
	\begin{IEEEeqnarray}{cCl}
		n R_1 &\leqslant& \sum_{t=1}^n I \left(X_{1,t};Y_{1,t}|X_{2,t}\right)+\epsilon_1^{(n)},\\
		n R_2 &\leqslant& \sum_{t=1}^n I \left(X_{2,t};Y_{1,t}|X_{1,t}\right)+\epsilon_2^{(n)},\\
		n(R_1+R_2) & \leqslant & \sum_{t=1}^n I\left(X_{1,t} X_{2,t};Y_{1,t}\right)+\epsilon_{12}^{(n)},
	\end{IEEEeqnarray}
\end{subequations}
where $\frac{\epsilon_1^{(n)}}{n},\frac{\epsilon_2^{(n)}}{n}$, and $\frac{\epsilon_1^{(n)}}{n}$ tend to zero as $n$ tends to infinity.

Using assumption~\eqref{assumption2}, for a given $\epsilon^{(n)}>0$, for any $\eta>0$ there exists $n_0(\eta)$ such that for any  $n\geq n_0(\eta)$ it holds that
\begin{equation}
\Pr\left(B^{(n)}<B-\epsilon^{(n)}\right) < \eta.
\end{equation}
Equivalently,
\begin{equation}
\Pr\left(B^{(n)}\geqslant B-\epsilon^{(n)}\right) \geqslant 1-\eta \label{eq:44}
\end{equation}

Using Markov's inequality~\cite{Durret-Book-2010}, the probability in \eqref{eq:44} can be upper-bounded as follows:
\begin{equation}
(B-\epsilon^{(n)})\Pr\left( B^{(n)} \geqslant B - \epsilon^{(n)} \right)  \leqslant \E{B^{(n)}}.\label{eq:45}
\end{equation}
Combining~\eqref{eq:44} and~\eqref{eq:45} yields   
\begin{equation}
(B-\epsilon^{(n)}) (1-\eta) \leqslant \E{B^{(n)}}
\end{equation}
which can be written as 
\begin{equation}
\label{eq:100}
(B-\delta^{(n)})\leqslant \E{B^{(n)}}
\end{equation}
for some $\delta^{(n)} >\epsilon^{(n)}$ (for sufficiently large $n$). Hence, \eqref{eq:sum_outer_bound} and \eqref{eq:100} are an upper-bound for any $(R_1,R_2,B)$ satisfying \eqref{assumption1} and \eqref{assumption2}.

In the following, the bounds in \eqref{eq:sum_outer_bound}, \eqref{eq:100}, and \eqref{assumption3} are evaluated for the \GMACF$(b)$. 
For this purpose, assume that $X_{1,t}$ and $X_{2,t}$ are arbitrary correlated random variables with  
\begin{IEEEeqnarray}{cCl}
	\mu_{i,t}&\eqdef &\E{X_{i,t}},\\
	\sigma_{i,t}^2&\eqdef& \Var{X_{i,t}},\\
	\lambda_t&\eqdef& \Cov{X_{1,t}}{X_{2,t}},
\end{IEEEeqnarray} 
for $t\in \{1,\dots,n\}$ and for $i\in \{1,2\}$. 

The input sequence must satisfy the input power constraint~\eqref{EqPowerConstraint} which can be written, for $i \in \{1,2\}$, as
\begin{equation}
\frac1n \sum_{t=1}^n \E{X_{i,t}^2}=\left(\frac1n \sum_{t=1}^n \sigma_{i,t}^2\right)+\left(\frac1n \sum_{t=1}^n\mu_{i,t}^2\right)\leqslant P_i.\label{eq:condpc1}
\end{equation} 
Note that from \eqref{EqY}, for each $t\in \{1,\dots,n\}$, it holds that
\begin{IEEEeqnarray}{rCl}
	h(Y_{1,t}|X_{1,t},X_{2,t})&=&h(Z_t)=\frac12 \log_2\left(2 \pi e \right),
\end{IEEEeqnarray}
from the assumption that $Z_t$ follows a zero-mean unit-variance Gaussian distribution. 
Note also that for any random variable $X$ with variance $\sigma_X^2$, it holds that  $h(X)\leqslant \frac12 \log_2\left(2 \pi e \sigma_X^2\right)$, with equality  when $X$ follows  a Gaussian distribution~\cite{EIT}. Finally, it is useful to highlight that  for any $a\in \Reals$, it holds that $h(X+a)=h(X)$. Using these elements, the right-hand side terms in \eqref{eq:sum_outer_bound} can be upper-bounded as follows:
\begin{IEEEeqnarray*}{rCl}
	I(X_{1,t}, X_{2,t};Y_{1,t})&=& h(Y_{1,t})-h(Z_t)\\
	&\leqslant&  \frac12 \log_2\left(2 \pi e \Var{Y_{1,t}}\right)-\frac12 \log_2\left(2 \pi e\right)\\
	&=&\frac12 \log_2\left(h_{11}^2 \sigma_{1,t}^2+h_{12}^2 \sigma_{2,t}^2+2 h_{11} h_{12} \lambda_t+1\right),\\
	I(X_{1,t};Y_{1,t}|X_{2,t})&=& h(Y_{1,t}|X_{2,t})-h(Y_{1,t}|X_{1,t},X_{2,t})\\
	&\leqslant & \frac12 \log_2\left(2 \pi e (\Var{Y_{1,t}|X_{2,t}})\right)\hspace*{-1mm}-\hspace*{-1mm}\frac12 \log_2\left(2 \pi e\right)\\
	&=&\frac12 \log_2\left(1+h_{11}^2\sigma_{1,t}^2\left(1-\frac{\lambda_t^2}{\sigma_{1,t}^2\sigma_{2,t}^2}\right)\right),\\
	I(X_{2,t};Y_{1,t}|X_{1,t})&=&\frac12 \log_2\left(1+h_{12}^2 \sigma_{2,t}^2\left(1-\frac{\lambda_t^2}{\sigma_{1,t}^2\sigma_{2,t}^2}\right)\right).
\end{IEEEeqnarray*}
Finally,  the bounds in~\eqref{eq:sum_outer_bound} can be rewritten as follows:
\begin{subequations}
	\label{eq:sum_outer_bound_gaussian}
	\begin{IEEEeqnarray}{cCl}
		\hspace*{-1mm}n R_1\hspace*{-1mm} &\leqslant&\hspace*{-1mm}\sum_{t=1}^n\hspace*{-1mm}\frac12\hspace*{-.8mm} \log_2\hspace*{-1mm}\left(\hspace*{-1mm}1\hspace*{-1mm}+\hspace*{-1mm}h_{11}^2 \sigma_{1,t}^2\hspace*{-1mm}\left(\hspace*{-1mm}1\hspace*{-1mm}-\hspace*{-1mm}\frac{\lambda_t^2}{\sigma_{1,t}^2\sigma_{2,t}^2}\hspace*{-1mm}\right)\hspace*{-1mm}\right)\hspace*{-1mm} +\hspace*{-1mm}\epsilon_1^{(n)}\hspace*{-1mm},\hspace*{-1mm}\IEEEeqnarraynumspace
		\\
		\hspace*{-1mm}	n R_2\hspace*{-1mm} &\leqslant&\hspace*{-1mm}\sum_{t=1}^n\hspace*{-1mm}\frac12 \hspace*{-.8mm}\log_2\hspace*{-1mm}\left(\hspace*{-1mm}1\hspace*{-1mm}+\hspace*{-1mm}h_{12}^2\sigma_{2,t}^2\hspace*{-1mm}\left(\hspace*{-1mm}1\hspace*{-1mm}-\hspace*{-1mm}\frac{\lambda_t^2}{\sigma_{1,t}^2\sigma_{2,t}^2}\hspace*{-1mm}\right)\hspace*{-1mm}\right)\hspace*{-1mm}+\hspace*{-1mm}\epsilon_2^{(n)}\hspace*{-1mm},
		\\
		\hspace*{-1mm}n (R_1+R_2)\hspace*{-1mm} & \leqslant &\hspace*{-1mm}\sum_{t=1}^n\hspace*{-1mm}\frac12\hspace*{-.5mm} \log_2\left(1\hspace*{-1mm}+\hspace*{-1mm}h_{11}^2 \sigma_{1,t}^2\hspace*{-1mm}+\hspace*{-1mm}h_{12}^2 \sigma_{2,t}^2
		\hspace*{-1mm}+\hspace*{-1mm}2 h_{11} h_{12} \lambda_t\right)+\epsilon_{12}^{(n)}.
	\end{IEEEeqnarray}
\end{subequations}

The expectation of the average received energy rate is given by
\begin{IEEEeqnarray}{rCl}
	\label{eq:142}
	\E{B^{(n)}} &=&\E{\frac{1}{n}\ds\sum_{t =1}^{n} Y_{2,t}^2 }  \nonumber\\&=&1+ h_{21}^2 \left(\hspace*{-1mm}\frac{1}{n}\ds\sum_{t =1}^n (\sigma_{1,t}^2+\mu_{1,t}^2) \hspace*{-1mm}\right)+h_{22}^2 \left(\hspace*{-1mm}\frac{1}{n}\ds\sum_{t =1}^n (\sigma_{2,t}^2+\mu_{2,t}^2)\right)\nonumber\\&&+2 h_{21} h_{22} \left(\frac{1}{n}\ds\sum_{t=1}^{n} (\lambda_t+\mu_{1,t} \mu_{2,t})\right).
\end{IEEEeqnarray}
Using the Cauchy-Schwarz inequality, the energy rate in~\eqref{eq:142} can be upper-bounded as follows:
\begin{IEEEeqnarray}{rCl}
	\E{B^{(n)}}\hspace*{-1mm}&\leqslant &\hspace*{-1mm}1\hspace*{-.5mm}+
	\hspace*{-1mm} h_{21}^2 \hspace*{-1mm}\left(\hspace*{-1mm}\frac{1}{n}\hspace*{-.8mm}\ds\sum_{t =1}^n (\sigma_{1,t}^2+\mu_{1,t}^2)\hspace*{-1mm}\right)\hspace*{-1mm}
	+\hspace*{-1mm}h_{22}^2\hspace*{-1mm}\left(\hspace*{-1mm}\frac{1}{n}\hspace*{-.8mm}\ds\sum_{t =1}^n (\sigma_{2,t}^2\hspace*{-1mm}+\hspace*{-1mm}\mu_{2,t}^2) \hspace*{-1mm}\right)\hspace*{-1mm}\nonumber\\&&\hspace*{-1mm}+2 h_{21} h_{22}\hspace*{-1mm} \left(\hspace*{-.5mm}\left|\hspace*{-1mm}\frac{1}{n}\hspace*{-.8mm}\ds\sum_{t=1}^{n}\hspace*{-.8mm} \lambda_t\right|\hspace*{-1mm}+\hspace*{-1mm}\left(\hspace*{-1mm}\frac{1}{n}\hspace*{-1mm}\sum_{t=1}^{n}\hspace*{-.6mm}\mu_{1,t}^2\hspace*{-1mm}\right)^{\hspace*{-1.5mm}1/2}\hspace*{-2mm} \left(\hspace*{-1mm}\frac{1}{n}\hspace*{-1mm}\sum_{t=1}^{n}\hspace*{-.6mm}\mu_{2,t}^2\hspace*{-1mm}\right)^{\hspace*{-1.5mm}1/2}\right)\hspace*{-1mm}.\label{eq:134}
\end{IEEEeqnarray}

Combining  \eqref{eq:100} and \eqref{eq:134} yields the following upper-bound on the energy rate $B$:
\begin{IEEEeqnarray}{rcl}
	\label{cstB}
	B&\leqslant & 1\hspace*{-.8mm}+\hspace*{-.8mm} h_{21}^2 \hspace*{-.8mm}\left(\hspace*{-.8mm}\frac{1}{n}\ds\sum_{t =1}^n (\sigma_{1,t}^2+\mu_{1,t}^2)\hspace*{-.8mm}\right)\hspace*{-.8mm}+\hspace*{-.8mm}h_{22}^2 \left(\hspace*{-.8mm}\frac{1}{n}\ds\sum_{t =1}^n (\sigma_{2,t}^2+\mu_{2,t}^2)\hspace*{-.8mm}\right)\hspace*{-.8mm}+\nonumber\\&&2 h_{21} h_{22} \hspace*{-1mm}\left(\hspace*{-.5mm}\left|\frac{1}{n}\ds\sum_{t=1}^{n} \lambda_t\right|\hspace*{-1mm}+\hspace*{-1mm}\left(\hspace*{-1mm}\frac{1}{n}\hspace*{-1mm}\sum_{t=1}^{n}\mu_{1,t}^2\right)^{\hspace*{-1.5mm}1/2} \hspace*{-2mm}\left(\hspace*{-1mm}\frac{1}{n}\hspace*{-1mm}\sum_{t=1}^{n}\mu_{2,t}^2\hspace*{-1mm}\right)^{\hspace*{-1.5mm}1/2}\right)\hspace*{-1mm}+\hspace*{-1mm}\delta^{(n)}\hspace*{-1mm}.
\end{IEEEeqnarray}

In order to obtain a single-letterization of the upper-bound given by constraints \eqref{eq:sum_outer_bound_gaussian} and \eqref{cstB}, define also 
\begin{IEEEeqnarray}{rCl}
	\mu_i^2&\eqdef& \frac{1}{n}\ds\sum_{t =1}^n \mu_{i,t}^2,\quad i \in \{1,2\},\\
	\sigma_i^2&\eqdef& \frac{1}{n}\ds\sum_{t =1}^n \sigma_{i,t}^2,\quad i \in \{1,2\},\\
	\rho&\eqdef&\left(\frac{1}{n}\ds\sum_{t=1}^{n} \lambda_t\right)\left(\left|\sigma_1\right|\left|\sigma_2\right|\right)^{-1}.
\end{IEEEeqnarray}

With these notations, the input power constraint in \eqref{eq:condpc1} can be rewritten as
\begin{equation}
\label{eq:146}
\sigma_{i}^2+\mu_{i}^2 \leqslant P_i,\quad i \in \{1,2\}.
\end{equation}

By the concavity of the logarithm, applying Jensen's inequality~\cite{EIT} in the bounds~\eqref{eq:sum_outer_bound_gaussian} yields, in the limit when $n\to \infty$,
\begin{subequations}
	\begin{IEEEeqnarray}{cCl}
		R_1&\leqslant& \frac12 \log_2\left(1+h_{11}^2 \sigma^2_1  \left(1-\rho^2\right)\right),\\
		R_2&\leqslant &\frac12 \log_2\left(1+h_{12}^2 \sigma_2^2 \left(1-\rho^2\right)\right),\\
		R_1+R_2&\leqslant& \frac12 \log_2\hspace*{-1mm}\left(\hspace*{-1mm}1+h_{11}^2 \sigma_1^2 + h_{12}^2 \sigma_2^2 +2 \sqrt{h_{11}^2 \sigma_1^2 h_{12}^2 \sigma_2^2}\rho\hspace*{-1mm}\right)\hspace*{-1mm},
	\end{IEEEeqnarray}
	and the upper-bound on the energy rate \eqref{cstB} yields
	\begin{IEEEeqnarray}{rCl}
		B &\leqslant & 1+ h_{21}^2 (\sigma_1^2+\mu_1^2)+h_{22}^2 (\sigma_2^2+ \mu_2^2) +2 h_{21} h_{22} \left(|\rho|\;|\sigma_1| |\sigma_2|+|\mu_1| |\mu_2|\right).
\end{IEEEeqnarray}\end{subequations}

Let $\set{R}_b(\sigma_1^2,\sigma_2^2,\mu_1,\mu_2,\rho)$ denote the set of  information-energy rate triplets $(R_1,R_2,B)$ satisfying:
\begin{subequations}
	\label{eq:147}
	\begin{IEEEeqnarray}{cCl}
		R_1&\leqslant&\hspace*{-1mm} \frac12 \log_2\left(1+h_{11}^2 \sigma^2_1  (1-\rho^2)\right),\\
		R_2&\leqslant &\hspace*{-1mm}\frac12 \log_2\left(1+h_{12}^2 \sigma_2^2 (1-\rho^2)\right),\\
		R_1+R_2\hspace*{-1mm}&\leqslant&\hspace*{-1mm} \frac12 \log_2\left(\hspace*{-0.5mm}1\hspace*{-0.5mm}+\hspace*{-0.5mm}h_{11}^2 \sigma_1^2\hspace*{-0.5mm} +\hspace*{-0.5mm} h_{12}^2 \sigma_2^2\hspace*{-0.5mm} +\hspace*{-0.5mm}2 \sqrt{h_{11}^2 h_{12}^2 \sigma_1^2\sigma_2^2}\rho\right),\\
		B&\leqslant&\hspace*{-1mm}1+ h_{21}^2 (\sigma_1^2+\mu_1^2)+h_{22}^2 (\sigma_2^2+ \mu_2^2) +2 h_{21} h_{22} (|\rho|\;|\sigma_1| |\sigma_2|+|\mu_1| |\mu_2|),\\
		B&\geqslant&b,
	\end{IEEEeqnarray}
\end{subequations}
for some $\sigma_1^2$, $\sigma_2^2$, $\mu_1$, $\mu_2$ such that~\eqref{eq:146} is true and for some $\rho\in [-1,1]$.  

To sum up, it has been shown so far that,
in the limit when $n$ tends to infinity, any information-energy rate triplet $(R_1,R_2,B)\in \set{E}_b^\FB(\SNR_{11},\SNR_{12},\SNR_{21},\SNR_{22})$ can be bounded by the constraints in \eqref{eq:147} for some $\sigma_1^2$, $\sigma_2^2$, $\mu_1$, $\mu_2$ satisfying~\eqref{eq:146} and for some $\rho\in [-1,1]$. Thus, it holds that
\begin{equation} \set{E}_b^\FB(\SNR_{11},\SNR_{12},\SNR_{21},\SNR_{22})\subseteq \bigcup_{\substack{	0\leqslant\sigma_1^2+\mu_1^2\leqslant P_1\\0\leqslant\sigma_2^2+\mu_2^2\leqslant P_2\\-1\leqslant \rho \leqslant 1}}
\set{R}_b(\sigma_1^2,\sigma_2^2,\mu_1,\mu_2,\rho).
\end{equation}

In this union, it suffices to consider ${0\leqslant \rho\leqslant 1}$ because for any ${-1\leqslant\rho\leqslant 1}$,\break ${\set{R}_b(\sigma_1^2,\sigma_2^2,	\mu_1^2,\mu_2^2,\rho)\subseteq\set{R}_b(\sigma_1^2,\sigma_2^2,\mu_1^2,\mu_2^2,|\rho|).}$ Furthermore, for $0\leqslant \rho \leqslant 1$, it suffices to consider $\mu_1\geqslant 0$, $\mu_2\geqslant 0$, and  $\sigma_1^2$, $\sigma_2^2$, $\mu_1^2$, and $\mu_2^2$ that saturate the input power constraint (i.e., \eqref{eq:146} holds with equality). Thus, 
\begin{multline} \set{E}_b^\FB(\SNR_{11},\SNR_{12},\SNR_{21},\SNR_{22})\\\subseteq \bigcup_{\substack{	0\leqslant\sigma_1^2+\mu_1^2\leqslant P_1\\0\leqslant\sigma_2^2+\mu_2^2\leqslant P_2\\-1\leqslant \rho \leqslant 1}}
\set{R}_b(\sigma_1^2,\sigma_2^2,\mu_1,\mu_2,\rho)\subseteq \bigcup_{\substack{	\sigma_1^2+\mu_1^2= P_1\\\sigma_2^2+\mu_2^2= P_2\\0\leqslant \rho \leqslant 1}}
\set{R}_b(\sigma_1^2,\sigma_2^2,\mu_1,\mu_2,\rho).
\end{multline}

Let $\beta_i\in [0,1]$ be defined as follows:
\begin{equation}
\beta_i\eqdef \frac{\sigma_i^2}{P_i}=\frac{P_i-\mu_i^2}{P_i}, \quad i\in\{1,2\}.
\end{equation}
With this notation, any region $\set{R}_b(\sigma_1^2,\sigma_2^2,\mu_1,\mu_2,\rho)$ in the union over $\sigma_1^2+\mu_1^2= P_1$, $\sigma_2^2+\mu_2^2= P_2$ and $0\leqslant \rho \leqslant 1$, can be rewritten as follows:
\begin{subequations}
	\begin{IEEEeqnarray}{cCl}
		R_1&\leqslant&\hspace*{-1mm} \frac12 \log_2\left(1+h_{11}^2 \beta_1 P_1 \left(1-\rho^2\right)\right),\\
		R_2&\leqslant &\hspace*{-1mm}\frac12 \log_2\left(1+h_{12}^2 \beta_2 P_2\left(1-\rho^2\right)\right),\\
		R_1+R_2\hspace*{-1mm}&\leqslant&\hspace*{-1mm} \frac12 \log_2\Bigr(\hspace*{-0.5mm}1\hspace*{-0.5mm}+\hspace*{-0.5mm}h_{11}^2 \beta_1 P_1\hspace*{-0.5mm} +\hspace*{-0.5mm} h_{12}^2 \beta_2 P_2\hspace*{-0.5mm}+\hspace*{-0.5mm}2 \sqrt{h_{11}^2 h_{12}^2 \beta_1 P_1 \beta_2 P_2}\rho\Bigr),\IEEEeqnarraynumspace\\
		B&\leqslant&\hspace*{-1mm}1+ h_{21}^2 P_1+h_{22}^2 P_2+2 h_{21} h_{22} (\rho\;\sqrt{\beta_1 P_1 \beta_2 P_2 }+\sqrt{(1-\beta_1)P_1 (1-\beta_2) P_2}),\\
		B&\geqslant&b,
	\end{IEEEeqnarray}
\end{subequations}
for some 
$(\beta_1,\beta_2)\in [0,1]^2$ and $\rho\in [0,1]$. Hence, using \eqref{eq:SNR}, such a region contains all information-energy rate triplets $(R_1,R_2,B)$  satisfying constraints \eqref{eq:regthm1}  which completes the proof of the converse.

\section{Proof of Theorem~\ref{Theorem-EC-Region-MAC-NF}}
\label{ProofTheorem-EC-Region-MAC-NF}
Consider that each transmitter $i$, with $i\in \{1,2\}$, uses a fraction $\beta_i\in [0,1]$ of its available power to transmit information and uses the remaining fraction of power $(1-\beta_i)$ to transmit energy.
Given a power-split $(\beta_1,\beta_2)\in [0,1]^2$, the achievability of information rate pairs satisfying \eqref{eq:thm1_c1}-\eqref{eq:thm1_c12} follows by the coding scheme proposed independently by Cover~\cite{COVER75} and Wyner~\cite{WYNER} with powers $\beta_1 P_1$ and $\beta_2 P_2$.
Additionally, in order to satisfy the received energy constraint~\eqref{eq:thm1_e}, transmitters send common randomness  that is known to both transmitters and the receiver using all their remaining power. This common randomness does not carry any information and does not produce any interference to the IC signals.
More specifically, at each time~$t$, transmitter~$i$'s  channel input can be written as:
\begin{IEEEeqnarray}{rCl}
	X_{i,t}=\sqrt{(1-\beta_i) P_i} W_t+U_{i,t},\quad i \in \{1,2\},
\end{IEEEeqnarray}
for some independent zero-mean Gaussian IC symbols
$U_{1,t}$ and $U_{2,t}$ with variances $\beta_1 P_1$ and $\beta_2 P_2$, respectively, and independent thereof $W_t$ is a zero-mean unit-variance Gaussian NIC symbol known non-causally to all terminals.

The receiver subtracts the common randomness and then performs successive decoding to recover the messages $M_1$ and $M_2$.  Note that this strategy achieves the corner points of the information rate-region at a given energy rate. Time-sharing between the corner points and the points on the axes is needed to achieve the remaining points.

The converse and the analysis of the average received energy rate follow along the lines of the case with feedback described in Appendix~\ref{SecProofThm3} when the IC channel input components are assumed to be independent.

\section{Proof of Proposition~\ref{PropMaxIndividualRate}}
\label{ProofPropMaxIndividualRate}
For a given energy transmission rate of $b$ energy-units per channel use, a power-split $(\beta_1,\beta_2)$ is feasible  if there exists at least one $\rho \in [0,1]$ that satisfies
\begin{equation}
g^\FB(\beta_i,\beta_j,\rho)\geqslant b,\label{EqFeasiblePowerSplits}
\end{equation}
with
\begin{IEEEeqnarray}{rCl}
	g^\FB(\beta_i,\beta_j,\rho)&\eqdef& 1 + \SNR_{2i} + \SNR_{2j} +2 \sqrt{(1-\beta_i)\SNR_{2i}(1-\beta_j)\SNR_{2j}}
	\nonumber\\&&+2 \rho \sqrt{\beta_i \SNR_{2i} \;\beta_j \SNR_{2j}}.\label{eq:grho}
\end{IEEEeqnarray}

Using a Fourier-Motzkin elimination in the constraints~\eqref{EqEC1}-\eqref{EqEC3} to eliminate $R_j$, it can be shown that transmitter $i$'s individual rate maximization problem~\eqref{EqRindiv} is equivalent to
\begin{subequations}	
	\label{probmaxrifb49}
	\begin{IEEEeqnarray}{lcl}
		R_i^{\FB}(b)&=& \max_{(\beta_i,\beta_j,\rho)\in [0,1]^3} f_i^{\FB}(\beta_i,\beta_j,\rho),\\
		\text{subject to:} & & g^{\FB}(\beta_i,\beta_j,\rho)\geqslant b,
	\end{IEEEeqnarray}
\end{subequations}
with
\begin{IEEEeqnarray}{l}
	f_i^\FB(\beta_i,\beta_j,\rho)\eqdef\min\Biggr\{\frac{1}{2} \log_2\left( 1 + \beta_i \;\SNR_{1i}\left( 1 - \rho^2 \right)\right),\nonumber\\\hspace*{2.5cm}\frac{1}{2}\log_2\hspace*{-1mm} \big( 1 +\beta_i\SNR_{1i} + \beta_j\SNR_{1j} + 2 \rho  \sqrt{\beta_i \SNR_{1i}\beta_j \SNR_{1j}}\big)\Biggr\}.
\end{IEEEeqnarray}

For a given triplet $(\beta_i,\beta_j,\rho)$, there are two cases: either it satisfies 
\begin{IEEEeqnarray}{l}
	-\rho^2 \beta_i \;\SNR_{1i}>\beta_j\SNR_{1j} + 2 \rho  \sqrt{\beta_i \SNR_{1i}\beta_j \SNR_{1j}},\IEEEeqnarraynumspace \label{cond53}
\end{IEEEeqnarray}
which implies that
\begin{IEEEeqnarray}{l}
	f_i^\FB(\beta_i,\beta_j,\rho)=\frac{1}{2}\hspace*{-1mm}\log_2\hspace*{-1mm}\big(1 +\beta_i\SNR_{1i} + \beta_j\SNR_{1j} + 2 \rho  \sqrt{\beta_i\SNR_{1i}\beta_j \SNR_{1j}}\big);
\end{IEEEeqnarray}
or it satisfies
\begin{IEEEeqnarray}{l}
	-\rho^2 \beta_i \;\SNR_{1i}\leqslant\beta_j\SNR_{1j} + 2 \rho  \sqrt{\beta_i \SNR_{1i}\beta_j \SNR_{1j}}, \label{eq:Cond1}\IEEEeqnarraynumspace
\end{IEEEeqnarray}
and in this case
\begin{IEEEeqnarray}{l}
	f_i^\FB(\beta_i,\beta_j,\rho)=\frac{1}{2} \log_2\left( 1 + \beta_i \;\SNR_{1i}\left( 1 - \rho^2 \right)\right).\IEEEeqnarraynumspace
\end{IEEEeqnarray}

In the first case, condition~\eqref{cond53} cannot be true for any triplet $(\beta_i,\beta_j,\rho)\in [0,1]^3$ and this case should be excluded. 

In the second case, the function $f_i^\FB(\beta_i,\beta_j,\rho)$ is decreasing in $\rho$ and does not depend on $\beta_j$, thus, it holds that 
\begin{equation}f_i^\FB(\beta_i,\beta_j,\rho)\leq f_i^\FB(\beta_i,0,0),
\end{equation}
and the triplet $(\beta_i,0,0)$ is feasible if and only if  $g^\FB(\beta_i,0,0)\geqslant b$. 
Under these assumptions, transmitter $i$ is able to achieve its maximum individual rate if
it uses a power-split in which the fraction $\beta_i$ is maximized and its 
energy transmission is made at the minimum
rate to meet the energy rate constraint.
In this case, the maximization problem \eqref{probmaxrifb49} reduces to the maximization problem in \eqref{probmaxindrNF} in the proof of Proposition~\ref{PropMaxIndividualRateNF}.
Thus, it can be shown that the individual rates with feedback are limited by $R_i \leqslant \frac{1}{2}\log_2\left( 1 + (1-\xi(b)^2)  \SNR_{1i} \right)$ where $\xi(b)$ is given by~\eqref{xi}.

\section{Proof of Proposition~\ref{PropMaxIndividualRateNF}}
\label{ProofPropMaxIndividualRateNF}
From the assumptions of Proposition~\ref{PropMaxIndividualRateNF} it follows that 
an energy transmission rate of $b$ energy-units per channel use must be guaranteed at the input of the EH. Then, the set of power-splits $(\beta_i,\beta_j)$ that satisfy this constraint must satisfy
\begin{equation}
g_0(\beta_i,\beta_j)
\geqslant b, 
\label{EqFeasiblePowerSplitsNF}
\end{equation}
with 
\begin{IEEEeqnarray}{rCl}
	\label{eq:g0}
	g_0(\beta_i,\beta_j)&\eqdef &1+\SNR_{21}+\SNR_{22}+ 2 \sqrt{(1-\beta_i)\SNR_{2i}(1-\beta_j)\SNR_{2j}}.\IEEEeqnarraynumspace
\end{IEEEeqnarray}	
These power-splits are referred to as \emph{feasible power-splits}. 

Using a Fourier-Motzkin elimination in the constraints~\eqref{eq:thm1_c1}-\eqref{eq:thm1_c12} to eliminate $R_j$, it can be shown that transmitter $i$'s individual rate maximization problem in 	\eqref{EqRindivNF} can be written as
\begin{subequations}
	\begin{IEEEeqnarray}{lcl}
		R_i^\NF(b)&=& \max_{(\beta_i,\beta_j)\in [0,1]^2} f_i(\beta_i,\beta_j),\\
		\text{subject to:}&&g_0(\beta_i,\beta_j)\geqslant b,
	\end{IEEEeqnarray}
\end{subequations}
with
\begin{IEEEeqnarray}{lcl}
	\label{eq:fi}
	f_i(\beta_i,\beta_j)&\eqdef&\min\biggr\{\frac12 \log_2\left(1+ \beta_i \SNR_{1i})\right),\frac12 \log_2 \left(1+ \beta_i \SNR_{1i}+\beta_j \SNR_{1j}\right)\biggr\}
\end{IEEEeqnarray}	
and $g_0(\beta_1,\beta_2)$ is defined in \eqref{eq:g0}.

For any feasible power-split $(\beta_i,\beta_j)$, it holds that 
\begin{equation}
f_i(\beta_i,\beta_j)= \frac12 \log_2 \left(1+ \beta_i \SNR_{1i}\right).
\end{equation}
The target function $f_i(\beta_i,\beta_j)$ is increasing in $\beta_i$ and is independent of $\beta_j$. Since the constraint function is monotonically decreasing in $(\beta_i,\beta_j)$, in order to maximize transmitter $i$'s individual rate,
the optimal power-split should be a feasible power-split in which $\beta_i$ is maximized while $\beta_j$ is forced to 0. Thus, the maximization problem in \eqref{EqRindivNF} can be written as follows:
\begin{subequations}
	\label{probmaxindrNF}
	\begin{IEEEeqnarray}{lcl}
		R_i^\NF(b)&=& \max_{\beta_i\in [0,1]} \frac12 \log_2(1+\beta_i \SNR_{1i}),\\
		\text{subject to:}&&g_0(\beta_i,0)\geqslant b.
	\end{IEEEeqnarray}
\end{subequations}
Transmitter $i$'s achievable information rate  is increasing in $\beta_i$ and the energy rate constraint is decreasing in $\beta_i$. 
Hence, transmitter $i$ is able to achieve the maximum individual rate if the energy transmission of transmitter $i$ is made at the minimum rate to meet the energy rate constraint, i.e., if there is equality in \eqref{EqFeasiblePowerSplitsNF}. In this configuration, transmitter $i$ can use a power-split in which $\beta_i=1-\xi(b)^2$, with $\xi(b)$ defined in \eqref{xi} which yields the maximum individual rate $R_i(b)= \frac{1}{2}\log_2\left( 1 + \left(1-\xi(b)^2\right)  \SNR_{1i} \right)$.

\section{Proof of Proposition~\ref{PropMaxSumRate}}
\label{SecProofThm4}
For fixed $\SNR_{11}$, $\SNR_{12}$, $\SNR_{21}$, and $\SNR_{22}$ and fixed minimum received energy rate $b\geqslant 0$ satisfying \eqref{EnergyCst}, the information sum-rate maximization problem in~\eqref{EqRsum} can be written as 
\begin{subequations}
	\label{eq:opt2}
	\begin{IEEEeqnarray}{lCl}
		R_{\mathrm{sum}}^\FB(b) &=&\max_{(\beta_1,\beta_2,\rho)\in [0,1]^3}
		f(\beta_1,\beta_2,\rho)\\ 
		\text{subject to:}\quad&\;& g(\beta_1,\beta_2,\rho)\geqslant b,\IEEEeqnarraynumspace
	\end{IEEEeqnarray}
\end{subequations}
where the functions $f$ and $g$ are defined as follows
\begin{IEEEeqnarray}{rCl}
	f(\beta_1,\beta_2,\rho)\eqdef\min\biggr\{ 
	&&\frac12\log_2\bigr(1+\beta_1 \SNR_{11}+\beta_2\SNR_{12}+ 2 \rho \sqrt{\beta_1 \SNR_{11} \beta_2 \SNR_{12} }\bigr),\nonumber\\
	&&\frac12\log_2\bigr(\left(1+\beta_1 \SNR_{11} (1-\rho^2)\right)\left(1+\beta_2 \SNR_{12} (1-\rho^2)\right)\bigr)\biggr\},
\end{IEEEeqnarray}
and
\begin{multline}
\hspace*{-3mm}g(\beta_1,\beta_2,\rho)\eqdef  1+\SNR_{21} +\SNR_{22}+2 \left(\sqrt{\beta_1 \beta_2} \rho+ \sqrt{(1-\beta_1) (1-\beta_2)}\right) \sqrt{\SNR_{21}\;\SNR_{22}}.
\end{multline}
Let also
\begin{IEEEeqnarray}{rCl}
		\rho_{\minr}(\beta_1,\beta_2)\eqdef\min\Biggr( 1,
	\hspace*{-1mm}\frac{\hspace*{-1mm}\left(\hspace*{-.8mm}b\hspace*{-1mm}-\hspace*{-1mm}\left(\hspace*{-1mm}1\hspace*{-0.8mm}+\hspace*{-0.8mm}\SNR_{21}\hspace*{-0.8mm}+\hspace*{-0.8mm}\SNR_{22}\hspace*{-0.8mm}+\hspace*{-0.8mm}2 \sqrt{\hspace*{-0.6mm}(1\hspace*{-0.8mm}-\hspace*{-0.8mm}\beta_1)\SNR_{21} (1\hspace*{-0.8mm}-\hspace*{-0.8mm}\beta_2)\hspace*{-0.6mm}\SNR_{22}\hspace*{-0.8mm} }\right)\hspace*{-1mm}\right)^{\hspace*{-1mm}+}}{2 \sqrt{\beta_1 \SNR_{21} \beta_2 \SNR_{22}}}\hspace*{-1mm} \Biggr)\hspace*{-1mm}
	\label{eq:rhomin}	
\end{IEEEeqnarray}
be the value of $\rho \in [0, 1]$ for which $g(\beta_1,\beta_2,\rho)=b$, with $\beta_1\neq 0$ and $\beta_2\neq 0$. Note that $\rho^\star(\beta_1,\beta_2)$, initially defined in \eqref{eq:rhostar},  can  be alternatively  defined as
\begin{IEEEeqnarray}{rCl}
	\label{eq:newrhostar}	
	\rho^\star(\beta_1,\beta_2)\eqdef\argmax_{\rho\in[0,1]}\quad f(\beta_1,\beta_2,\rho).
\end{IEEEeqnarray}	
when $\beta_1\neq 0$ and $\beta_2\neq 0$. When either $\beta_1=0$ or $\beta_2=0$ then $\rho^\star(\beta_1,\beta_2)=0$.

Using this notation, the proof of Proposition~\ref{PropMaxSumRate} is based on the following two lemmas.

\begin{lemma}\label{lem2}
	Let $(\beta_1,\beta_2, \rho)\in [0,1]^3$ be a solution to \eqref{eq:opt2}. Then,
	\begin{equation}
	\rho=\max\bigr\{\rho_\minr(\beta_1,\beta_2), \rho^\star(\beta_1,\beta_2)\bigr\}.
	\end{equation}	
\end{lemma}
\begin{IEEEproof}
	
	Let $(\beta_1,\beta_2)\in (0,1]^2$ be fixed. A necessary condition for $(\beta_1,\beta_2,\rho)$ to be feasible, i.e, $g(\beta_1,\beta_2,\rho)\geqslant b$, is $\rho \in [\rho_\minr(\beta_1,\beta_2),1]$, with $\rho_\minr(\beta_1,\beta_2)$ defined in \eqref{eq:rhomin}.
	
	Let $\bar{\rho}(\beta_1,\beta_2)$ be the solution to the following optimization problem:
	\begin{IEEEeqnarray}{rCl}
		\max_{\rho \in [\rho_\minr(\beta_1,\beta_2),1]} f(\beta_1,\beta_2,\rho).
	\end{IEEEeqnarray} 
	
	Assume that 
	\begin{equation}
	\label{EqRhominrhostar}
	\rho_\minr(\beta_1,\beta_2)\leqslant\rho^\star(\beta_1,\beta_2).
	\end{equation} 
	
	In this case, it follows that $(\beta_1,\beta_2,\rho^\star(\beta_1,\beta_2))$ is feasible.  
	
	From \eqref{eq:newrhostar}, it holds that $\forall \rho \in [\rho_\minr(\beta_1,\beta_2),1],$
	\begin{equation}
	f(\beta_1,\beta_2,\rho)\leqslant f(\beta_1,\beta_2,\rho^\star(\beta_1,\beta_2)).
	\end{equation}

	Hence, under condition \eqref{EqRhominrhostar}, $\bar\rho(\beta_1,\beta_2)=\rho^\star(\beta_1,\beta_2)$.
	
	Assume now that 
	\begin{equation}
	\rho_\minr(\beta_1,\beta_2)>\rho^\star(\beta_1,\beta_2).
	\end{equation} 
	
	In this case, for any $\rho \in  [\rho_\minr(\beta_1,\beta_2),1]$, it holds that
	\begin{IEEEeqnarray}{lcl}
		f(\beta_1,\beta_2,\rho)&=&\frac12 \log_2\left(1+\beta_1\SNR_{11} (1-\rho^2)\right)+\frac12 \log_2\left(1+\beta_2\SNR_{12} (1-\rho^2)\right).
	\end{IEEEeqnarray}
	Hence, $f$ is monotonically decreasing in $\rho$, and  thus $\bar\rho(\beta_1,\beta_2)=\rho_\minr(\beta_1,\beta_2)$. 

	Given that the statements above hold for any  pair $(\beta_1,\beta_2)$, then for any solution $(\beta_1,\beta_2,\rho)$ to \eqref{eq:opt2}, it follows that $\rho=\bar{\rho}(\beta_1,\beta_2)$. This completes the proof.
\end{IEEEproof}

\begin{lemma} \label{lem3}
	The unique solution to \eqref{eq:opt2} in $[0,1]^3$ is $(1,1,\bar{\rho})$ with 		
	\begin{equation}
	\label{Eq}
	\bar{\rho}\eqdef \max \bigr\{\rho_\minr(1,1), \rho^\star(1,1)\bigr\}.
	\end{equation}
\end{lemma}
\begin{IEEEproof}
	Assume that there exists another solution $(\beta'_1,\beta'_2,\rho')$ to \eqref{eq:opt2} different from $(1,1,\bar{\rho})$.
	Thus, for any $(\beta_1,\beta_2,\rho)\in [0,1]^3$ it holds that
	\begin{equation}
	f(\beta_1,\beta_2,\rho) \leqslant f(\beta'_1,\beta'_2,\rho').
	\end{equation}
	Note that for a fixed $\rho' \in [0,1]$,  $f(\beta_1,\beta_2,\rho')$ is strictly increasing in $(\beta_1,\beta_2)$. Hence, for any $(\beta_1,\beta_2)\in [0,1)^2$,  
	\begin{IEEEeqnarray}{rCl}
		\label{EqVerification}
		f(\beta_1,\beta_2,\rho')& < & f(1,1,\rho')\\&\leqslant &f(1,1,\bar{\rho}),
	\end{IEEEeqnarray} 
	where the second inequality follows by Lemma~\ref{lem2}. 
	Moreover, since $\bar{\rho} \geqslant \rho_\minr(1,1)$, the following inequality also holds:
	\begin{equation}
	g(1,1,\bar{\rho})\geqslant b.
	\end{equation}
	In particular, if $(\beta_1,\beta_2)=(\beta'_1,\beta'_2)$ in \eqref{EqVerification}, it follows that
	\begin{equation}
	f(\beta'_1,\beta'_2,\rho') <  f(1,1,\bar{\rho}),
	\end{equation}
	which contradicts the initial assumption that there exists a solution other than $(1,1,\bar{\rho})$. This establishes a proof of Lemma~\ref{lem3}.
\end{IEEEproof}	
Finally, the proof of Proposition~\ref{PropMaxSumRate} follows from the following equality:
\begin{equation}
\label{EqFinalRsum}
R_{\mathrm{sum}}^\FB(b) = f(1,1,\bar{\rho}).
\end{equation}
Note that when $b\in [0,1+\SNR_{21}+\SNR_{22}+2 \rho^\star(1,1) \sqrt{\SNR_{21}\SNR_{22}}]$, $\bar{\rho}=\rho^\star(1,1)$ and 
\begin{IEEEeqnarray}{lcl}
		R_{\mathrm{sum}}^\FB(b)=\frac12\hspace*{-.5mm} \log_2\hspace*{-1mm}\left(\hspace*{-.8mm}1\hspace*{-1mm}+\hspace*{-1mm}\SNR_{11}\hspace*{-1mm}+\hspace*{-1mm}\SNR_{12}\hspace*{-1mm}+\hspace*{-1mm}2 \rho^\star(1,1)\hspace*{-.5mm} \sqrt{\hspace*{-.8mm}\SNR_{11}\SNR_{12}}\right)\hspace*{-1mm}.\IEEEeqnarraynumspace
\end{IEEEeqnarray}
When $b\in [1+\SNR_{21}+\SNR_{22}+2 \rho^\star(1,1) \sqrt{\SNR_{21}\SNR_{22}}, 1+\SNR_{21}+\SNR_{22}+2 \sqrt{\SNR_{21}\SNR_{22}}]$, $\bar{\rho}=\rho_\minr(1,1)$ and 
\begin{IEEEeqnarray}{lcl}
	R_{\mathrm{sum}}^\FB(b)&=&\frac{1}{2} \log_2(1 +  (1-\xi(b)^2) \SNR_{11})+ \frac{1}{2}\log_2(1 + (1-\xi(b)^2) \SNR_{12}),
\end{IEEEeqnarray}
and this completes the proof.

\section{Proof of Proposition~\ref{PropMaxSumRateNF}}
\label{SecProofProp2}

The sum-rate maximization problem in~\eqref{EqRsumNF} can be written as follows:
\begin{subequations}
	\label{eq:powerSplittingNoFb}
	\begin{IEEEeqnarray}{rCl}
		R_{\mathrm{sum}}^\NF(b) &=&\max_{(\beta_1,\beta_2)\in [0,1]^2} f_0(\beta_1,\beta_2)\hspace*{8mm}\\
		\text{subject to:}&& g_0(\beta_1,\beta_2)\geqslant b,
	\end{IEEEeqnarray}		
\end{subequations}
where the functions $f_0$ and $g_0$ are defined as
\begin{IEEEeqnarray}{lcl}
		f_0(\beta_1,\beta_2)\eqdef && \min\biggr\{\frac12 \log_2(1+\beta_1 \SNR_{11}+\beta_2 \SNR_{12}),\nonumber\\&&\frac12 \log_2(1+\beta_1 \SNR_{11})+\frac12 \log_2(1+\beta_2 \SNR_{12})\biggr\}\label{eq:f0}	
\end{IEEEeqnarray}
and $g_0(\beta_1,\beta_2)$ defined as in \eqref{eq:g0}.

For any nonnegative  $\beta_1$ and  $\beta_2$ it can be shown that
\begin{equation}
f_0(\beta_1,\beta_2)=\frac12 \log_2(1+\beta_1 \SNR_{11}+\beta_2 \SNR_{12}),
\end{equation}
and thus the function $f_0$ is monotonically increasing in $(\beta_1,\beta_2)$. 
The function $g_0$ is monotonically decreasing in $(\beta_1,\beta_2)$.

\begin{lemma}
	\label{LemEqPC}
	A necessary condition for $(\beta_1^*,\beta_2^*)$ to be a solution to the optimization problem in \eqref{eq:powerSplittingNoFb} is to satisfy 
	\begin{equation}
	g_0(\beta_1^*,\beta_2^*)=b,
	\end{equation} 
	when $1+\SNR_{21}+\SNR_{22}<b\leqslant 1+\SNR_{21}+\SNR_{22}+2 \sqrt{\SNR_{21}\SNR_{22}}$, and 
	\begin{equation}
	\beta_1^*=\beta_2^*=1
	\end{equation}
	when $0\leqslant b\leqslant 1 + \SNR_{21}+\SNR_{22}$. 
\end{lemma}

\begin{IEEEproof}
	Let $(\beta_1^*,\beta_2^*)$ be a solution to the optimization problem in  \eqref{eq:powerSplittingNoFb}.
	
	Assume that $1+\SNR_{21}+\SNR_{22}<b\leqslant 1+\SNR_{21}+\SNR_{22}+2 \sqrt{\SNR_{21}\SNR_{22}}$ and $g_0(\beta_1^*,\beta_2^*)>b$. Without loss of generality, consider transmitter $1$. Since $g_0$ is monotonically decreasing in $\beta_1$ whereas $f_0$ is monotonically increasing in $\beta_1$,  there always exists a $\beta_1>\beta_1^*$ such that  $g_0(\beta_1,\beta_2^*)=b$ and $f_0(\beta_1,\beta_2^*)>f_0(\beta_1^*,\beta_2^*)$, which contradicts the assumption of the lemma.
	
	Assume $0\leqslant b\leqslant 1+\SNR_{21}+\SNR_{22}$ and assume without loss of generality that transmitter 1 uses a power-split $\beta_1^*<1$. From the initial assumption, the pair $(1,\beta_2^*)$ satisfies $g_0(1,\beta_2^*)\geq b$ and $f_0(1,\beta_2^*)\geqslant f_0(\beta_1^*,\beta_2^*)$ which contradicts the assumption of the lemma and completes the proof. 
\end{IEEEproof}	

From Lemma \ref{LemEqPC}, the optimization problem in \eqref{eq:powerSplittingNoFb} is equivalent to
\begin{subequations}
	\begin{IEEEeqnarray}{rCl}
		R_{\mathrm{sum}}^\NF(b) &=&\max_{(\beta_1,\beta_2)\in [0,1]^2} f_0(\beta_1,\beta_2)\hspace*{8mm}\\
		\text{subject to:}&& g_0(\beta_1,\beta_2)= b,
	\end{IEEEeqnarray}		
\end{subequations}

Assume that $0 \leqslant b \leqslant 1+\SNR_{21}+\SNR_{22}$. Then, from Lemma \ref{LemEqPC}, it follows that the solution to the optimization problem in \eqref{eq:powerSplittingNoFb} is $\beta_1^* = \beta_2^* = 1$.  

Assume now that
\begin{multline}
\label{Conditionb}
\hspace*{-.5cm}	1+\SNR_{21}+\SNR_{22}<b\leqslant 1+\SNR_{21}+\SNR_{22}+2 \sqrt{\SNR_{21}\SNR_{22}} \min\left\{ \sqrt{\frac{\SNR_{12}}{\SNR_{11}}}, \sqrt{\frac{\SNR_{11}}{\SNR_{12}}} \right\}.
\end{multline}

Note that for any energy rate constraint $b$ satisfying \eqref{Conditionb}, it holds that
\begin{equation}
\label{ConditionXi}
0<\xi(b)\leqslant  \min\left\{\sqrt{\frac{\SNR_{11}}{\SNR_{12}}},\sqrt{\frac{\SNR_{12}}{\SNR_{11}}}\right\}.
\end{equation}

Let $(\beta_1^*,\beta_2^*)$ be a feasible pair, i.e., $g_0(\beta_1^*,\beta_2^*)=b$. This can be rewritten in terms of $\xi(b)$ as follows:
\begin{equation}
\label{cond38}
(1-\beta_1^*)(1-\beta_2^*)=\xi(b)^2,
\end{equation}
with $\xi(b)$ defined in \eqref{xi}.

Note also that any solution to \eqref{cond38}, must satisfy that $\beta_1  \leqslant 1-\xi(b)^2$ and $\beta_2  \leqslant 1-\xi(b)^2$. 
Hence, to obtain the solution of the optimization problem in \eqref{eq:powerSplittingNoFb}, it suffices to perform the maximization  over all $(\beta_1,\beta_2)\in [0,1-\xi(b)^2]$.

Let $\beta_2^*\in [0,1-\xi(b)^2]$ be fixed. Then,  there is a unique feasible  choice of $\beta_1^*$ to satisfy \eqref{cond38}, given by
\begin{equation}
\beta_1^*=1-\frac{\xi(b)^2}{1-\beta_2^*}.
\end{equation}
The corresponding sum-rate is given by
\begin{IEEEeqnarray}{rcl}
	\kappa(\beta_2^*)&\eqdef& f_0(\beta_1^*,\beta_2^*)
= \frac12 \log_2\hspace*{-1mm}\left(\hspace*{-1mm}1\hspace*{-1mm}+\hspace*{-1mm}\left(\hspace*{-1mm}1-\frac{\xi(b)^2}{1-\beta_2^*}\right)\hspace*{-1mm} \SNR_{11}\hspace*{-1mm}+\hspace*{-1mm}\beta_2^* \SNR_{12}\hspace*{-1mm}\right)\hspace*{-1mm},\IEEEeqnarraynumspace
\end{IEEEeqnarray}
which is a concave function of $\beta_2^*$. Hence, given a fixed $\beta_1^*$, the unique optimal $\beta_2^*$ must be a solution to   $\frac{\mathrm{d} \kappa (\beta_2^*) }{\mathrm{d} \beta_2^*}= 0$. That is, 
\begin{IEEEeqnarray}{rcl}
	\label{EqDerivative0}
	(1-\beta_2^*)^2=\xi(b)^2 \frac{\SNR_{11}}{\SNR_{12}}.
\end{IEEEeqnarray} 	
The equality in \eqref{EqDerivative0} admits a solution in $[0,1-\xi(b)^2]$ if and only if \eqref{ConditionXi} is satisfied. This unique solution is given by
\begin{IEEEeqnarray}{rcl}
	\bar{\beta}_2^*=1-\xi(b) \sqrt{\frac{\SNR_{11}}{\SNR_{12}}}
\end{IEEEeqnarray}
and the corresponding $\bar{\beta}_1^*$ is given by
\begin{IEEEeqnarray}{rcl}
	\bar{\beta}_1^*=1-\xi(b) \sqrt{\frac{\SNR_{12}}{\SNR_{11}}}.
\end{IEEEeqnarray}
In this case, the sum-rate is
\begin{IEEEeqnarray}{rcl}
	\bar{R}_s&=&f_0(\bar{\beta}_1^*,\bar{\beta}_2^*)=\frac12\hspace*{-.5mm} \log_2\hspace*{-.5mm}\left(\hspace*{-.5mm}1\hspace*{-.5mm}+\hspace*{-.5mm}\SNR_{11}\hspace*{-.5mm}+\hspace*{-.5mm}\SNR_{12}\hspace*{-.5mm}-\hspace*{-.5mm} 2 \xi(b)\hspace*{-.5mm}  \sqrt{\hspace*{-.5mm}\SNR_{11}\SNR_{12}}\hspace*{-.5mm}\right)\hspace*{-1mm}.\IEEEeqnarraynumspace
\end{IEEEeqnarray}

Assume now that
\begin{multline}
\label{Conditionb1}
\hspace*{-3mm}1\hspace*{-.5mm}+\hspace*{-.5mm}\SNR_{21}\hspace*{-.5mm}+\hspace*{-.5mm}\SNR_{22}\hspace*{-.5mm}+\hspace*{-.5mm}2\hspace*{-.5mm}\sqrt{\hspace*{-.5mm}\SNR_{21}\SNR_{22}}\hspace*{-.5mm} \min\hspace*{-1mm}\left\{\hspace*{-1mm}\sqrt{\hspace*{-.5mm}\frac{\SNR_{12}}{\SNR_{11}}},\hspace*{-.5mm} \sqrt{\hspace*{-.5mm}\frac{\SNR_{11}}{\SNR_{12}}}\hspace*{-.5mm}\right\}\\<b\leqslant 1+\SNR_{21}+\SNR_{22}+2 \sqrt{\SNR_{21}\SNR_{22}}.
\end{multline}
This is equivalent to
\begin{equation}
\min\left\{ \sqrt{\frac{\SNR_{12}}{\SNR_{11}}}, \sqrt{\frac{\SNR_{11}}{\SNR_{12}}} \right\}\leqslant \xi(b)\leqslant 1.
\end{equation}
Under this condition, the only feasible pairs, i.e., solutions to  $g_0(\beta_1,\beta_2)=b$, are $(0,1-\xi(b)^2)$ and $(1-\xi(b)^2,0)$. Hence, for all $i \in \{1,2\}$ satisfying $i=\ds \argmax_{k\in \{1,2\}} \SNR_{1,k}$ and $j \in \{1,2\}\setminus \{i\}$, it follows that the solution to \eqref{eq:powerSplittingNoFb} is given by $\beta_i^*=1-\xi(b)^2$ and $\beta_j^*=0$ and this completes the proof.

\section{Proof of Theorem~\ref{LemmaEnergyRateAtSCWF}}
\label{ProofLemmaEnergyRateAtSCWF}
From Proposition~\ref{PropMaxSumRate}, for any $B\in [0,1+\SNR_{21}+\SNR_{22}+2 \rho^\star(1,1) \sqrt{\SNR_{21} \SNR_{22}}]$, $R_{\mathrm{sum}}^\FB(B)>R_{\mathrm{sum}}^\NF(0)$, and thus any $B\in [0,1+\SNR_{21}+\SNR_{22}+2 \rho^\star(1,1) \sqrt{\SNR_{21} \SNR_{22}}]$ cannot be a solution to the optimization problem in \eqref{EqBFB}. Hence, a necessary condition for 
$B$ to be a solution to the optimization problem in \eqref{EqBFB} is to satisfy $B \in (1+\SNR_{21}+\SNR_{22} + 2\rho^{\star}(1,1)\sqrt{\SNR_{21} \SNR_{22}},1+\SNR_{21}+\SNR_{22} + 2 \sqrt{\SNR_{21} \SNR_{22}}]$. Thus, from Proposition~\ref{PropMaxSumRate},  
the optimization problem in \eqref{EqBFB} can be rewritten as follows:
\begin{IEEEeqnarray}{l}
	\nonumber
	B_{\F} = \ds\max_{B \in (b_1, b_2] }   B \\
	\nonumber
	\mathrm{subject \; to: }\\ 
	\frac{1}{2}\log_2\hspace*{-.5mm}\left(\hspace*{-.5mm}1 \hspace*{-.5mm}+\hspace*{-.5mm} \SNR_{11}\hspace*{-.5mm} +\hspace*{-.5mm} \SNR_{12} \hspace*{-.5mm} \right)\hspace*{-.5mm}  = \hspace*{-.5mm}\frac{1}{2}\hspace*{-.5mm}\log_2\hspace*{-.5mm}\left(\hspace*{-.5mm} 1 \hspace*{-.5mm}+\hspace*{-.5mm} \left(\hspace*{-.5mm}1\hspace*{-.5mm}-\hspace*{-.5mm}\xi(B)^2\hspace*{-.5mm}\right) \SNR_{11}\right)\hspace*{-.5mm}  
	\label{EqOPmaxEnergyRate2}
	+\hspace*{-.5mm}\frac{1}{2}\log_2\left(\hspace*{-.5mm} 1 \hspace*{-.5mm}+\hspace*{-.5mm} \left(1-\xi(B)^2\right) \SNR_{12} \hspace*{-.5mm}\right).\IEEEeqnarraynumspace
\end{IEEEeqnarray}
where $b_1 = 1 + \SNR_{21} + \SNR_{22} + 2\rho^{\star}(1,1) \sqrt{\SNR_{21} \SNR_{22}}$ and $b_2 = 1+\SNR_{21}+\SNR_{22} + 2 \sqrt{\SNR_{21} \SNR_{22}}$.
The constraint of the problem \eqref{EqOPmaxEnergyRate2} induces a unique value for $\left(1-\xi(B)^2\right)$ within $[0,1]$ for each $B$, and thus, the optimization is vacuous. This implies that the unique solution $B_{\F}$ satisfies 
\begin{IEEEeqnarray}{ll}
	\left(1-\xi(B_{\F})^2\right)\hspace*{-1mm}=  
	& \frac{\SNR_{11} \hspace*{-1mm}+\hspace*{-1mm} \SNR_{12}}{2\SNR_{11}\SNR_{12}}\hspace*{-1mm}\left[\hspace*{-.8mm}\sqrt{ \hspace*{-.8mm}1\hspace*{-1mm}+\hspace*{-1mm} \frac{4 \SNR_{11} \SNR_{12}}{\SNR_{11}+\SNR_{12}}}\hspace*{-.5mm}-\hspace*{-.5mm}1\right]\hspace*{-.5mm}.
	\label{EqBetaStarBF}
\end{IEEEeqnarray}
Following the definition of $\xi$ in \eqref{xi} and solving for $B_{\F}$ in \eqref{EqBetaStarBF} yields \eqref{EqBLemma}. This completes the proof of Theorem~\ref{LemmaEnergyRateAtSCWF}.

\section*{Acknowledgments}
The authors would like to thank Prof. Osvaldo Simeone for his insightful comments. The authors gratefully acknowledge the reviewers for their careful reading and their suggestions.

\balance

\end{document}